\documentclass[
 aip,
 amsmath,amssymb,
 preprint,
]{revtex4-2}

\usepackage{graphicx}
\usepackage{dcolumn}
\usepackage{bm}
\usepackage{xcolor}
\usepackage{amsmath}
\usepackage{amssymb}
\usepackage[utf8]{inputenc}
\usepackage[T1]{fontenc}
\usepackage{mathptmx}
\usepackage{etoolbox}

\makeatletter
\def\@email#1#2{
 \endgroup
 \patchcmd{\titleblock@produce}
  {\frontmatter@RRAPformat}
  {\frontmatter@RRAPformat{\produce@RRAP{*#1\href{mailto:#2}{#2}}}\frontmatter@RRAPformat}
  {}{}
}

\makeatother
\begin{document}

\title[]{
Exploring Bifurcations in Bose-Einstein Condensates via Phase Field Crystal Models}

\author{A. B. Steinberg}
\altaffiliation{ORCID: 0000-0001-6598-9700}
\affiliation{Institut für Theoretische Physik, Westfälische Wilhelms-Universität Münster, Wilhelm-Klemm-Strasse 9, 48149 Münster, Germany}

\author{F. Maucher}
\altaffiliation{ORCID: 0000-0002-5808-3967}
\affiliation{Departament de Física, Universitat de les Illes Balears \& IAC-3, Campus UIB, E-07122 Palma de Mallorca, Spain}

\author{S. V. Gurevich}
\altaffiliation{ORCID: 0000-0002-5101-4686}

\author{U. Thiele}
\altaffiliation{ORCID: 0000-0001-7989-9271; $\quad$ Electronic mail: u.thiele@uni-muenster.de}
\homepage{http://www.uwethiele.de}
\affiliation{Institut für Theoretische Physik, Westfälische Wilhelms-Universität Münster, Wilhelm-Klemm-Strasse 9, 48149 Münster, Germany}
\affiliation{Center of Nonlinear Science (CeNoS), Westfälische Wilhelms-Universität Münster, Corrensstrasse 2, 48149 Münster, Germany}

\date{\today}

\begin{abstract}
To facilitate the analysis of pattern formation and of the related phase transitions in Bose-Einstein condensates (BECs) we present an explicit approximate mapping from the nonlocal Gross-Pitaevskii equation with cubic nonlinearity to a phase field crystal (PFC) model. This approximation is valid close to the superfluid-supersolid phase transition boundary. The simplified PFC model permits the exploration of bifurcations and phase transitions via numerical path continuation employing standard software. While revealing the detailed structure of the bifurcations present in the system, we demonstrate the existence of localized states in the PFC approximation. Finally, we discuss how higher-order nonlinearities change the structure of the bifurcation diagram representing the transitions found in the system.
\end{abstract}

\maketitle
\begin{quotation}
The formation of patterns in Bose-Einstein condensates (BECs) has been of great interest in recent years, but their description through nonlinear integro-differential equations makes detailed analyses tedious. An approximation in the vicinity of the superfluid-supersolid phase transition boundary in the form of a phase field crystal (PFC) model allows us to investigate the bifurcation structure and existence regions of different patterns. Of particular interest are localized states, which are commonly investigated for PFC-type models, but have rarely been discussed for Bose-Einstein condensates. Furthermore, we discuss how high-order nonlinearities affect the structure of the underlying transitions. 
\end{quotation}

\section{\label{sec:1}Introduction}

The formation of patterns and structures has always captivated scientists from a broad range of physical disciplines \cite{Pism2006springer,CrHo1993rmp} with topics reaching from the dendritic structure of snowflakes \cite{Ball2011n} to patterns of well defined length scales in reaction-diffusion systems or in morphogenesis.~\cite{Turi1952ptrslb} A particularly universal model to describe formation of regular patterns in various systems is the Swift-Hohenberg equation.~\cite{SwHo1977pra} Originally, it was formulated for Rayleigh–B{\'e}nard convection but also arises in a number of chemical,~\cite{HiDB1996pla} ecological~\cite{LBCL2009jtb} and optical~\cite{TlML1994prl} systems, as well as for elastic materials~\cite{SLTR2015nm} to name just a few. 

Five decades ago speculations on a quantum state of matter featuring discrete translational symmetry whilst maintaining a large superfluid fraction brought together traditional pattern formation and quantum physics communities.~\cite{AnLi1969spjetp,Legg1970prl, Ches1970pra} These states of matter are referred to as supersolids. 
At first, the realisation of such a phase remained elusive. However, recent pioneering experiments~\cite{KSWW2016nature,SWBF2016nature} involving dipolar Bose-Einstein condensates (BECs) revealed that beyond-mean field contributions describing quantum-fluctuations can lead to the stabilization of a supersolid phase, a fact that is also true for mixtures of BECs.~\cite{Petr2015prl,CTSN2018science} 
Ultimately, ultracold dipolar quantum gases turned out to be excellent experimental platform to explore supersolidity.~\cite{KSWW2016nature,SWBF2016nature,TLFC2019prl,BSWH2019prx,CPIN2019prx,TRLF2019nature,GBHS2019nature,NBPP2019prl,NPKP2021nature,TMBF2021science} 
Here, we take a step back for simplicity and to furthermore complement the discussions on the corresponding dipolar system. In particular, we analyze a commonly adopted prototypical toy model of soft-core bosons,~\cite{HeNP2010prl,Cinti:PRL:2010,SaMB2012prl,MMCP2013pra,MaSC2014jltp,CiBP2014njp} which approximates the interaction that can be realized using the so-called Rydberg-blockade effect~\cite{LFCD2001prl,HeNP2010prl,MHSK2011prl}. Here, we primarily seek to present the method rather than to connect it to specific physical systems.

In order to include supersolidity into the frame of the well-established field of pattern formation we first derive an approximate mapping valid at the superfluid-supersolid phase transition of the nonlocal Gross-Pitaevskii equation (GPE) to a so-called conserved Swift-Hohenberg (SH) equation, also referred to as phase field crystal (PFC) model,\cite{ElGr2004aps,ELWG2012aip,TARG2013pre} which only involves local nonlinearities (cf.~Sec.~\ref{sec:2}). 
This mapping is sufficiently general to be applicable to other settings involving nonlocal nonlinearities. For example, there are a range of optical systems that are also governed by nonlocal nonlinearities and are suspected to give rise to pattern formation.~\cite{SHAP2011prl,MPSK2016prl} 
This basic idea has already been formulated and explored,~\cite{HeBD2019pra} however, here we propose an alternative mapping that aims at approximating the behavior of the nonlocal GPE close to the transition to modulated states. 

Our approach permits, on the one hand, the employment of well-established methods for numerical path continuation~\cite{KrOG2007springer,EGUW2019springer} and, on the other hand, for potential synergies of concepts. One such synergy is related to the concept of localized states that thus far has not yet been explored in depth in the context of BECs. Localized states refer to a situation in which a finite patch of one pattern coexists with a background of another state, often a uniform one.\cite{Knob2015arcmp} In the context of phase transitions they occur in parameter regions of phase coexistence.~\cite{TFEK2019njp,HAGK2021ijam} In extended parameter ranges localized states display distinctive features that occur for many individual physical systems. A prominent example are bifurcation diagrams featuring a snakes-and-ladders structure, also referred to as 'homoclinic snaking'.\cite{BuKn2006pre,WoCh1999pdnp,Knob2015arcmp} Closer to critical points branches might only 'wiggle' or even be quite straight, depending on parameter values and considered ensemble.\cite{TARG2013pre,HAGK2021ijam,Knob2016ijam}

Employing numerical path continuation using, e.g., the standard package \textsc{pde2path},\cite{UeWR2014cup} permits one to trace out unstable states as well as stable ones. This allows for a comprehensive exploration of the structure of bifurcation diagrams. The method is most readily applied to local formulations, i.e.\ partial differential equations. Its numerical complexity makes it challenging to apply it to integro-differential equations like the nonlocal GPE. Therefore, here we employ continuation for the local approximation and explore the advantages and limitations of this approach. Furthermore, to limit computational effort we restrict our considerations to one and two spatial dimensions. Practically, this can be achieved via tight trapping and has been employed e.g. to suppress instabilities~\cite{Koch2008} or the dimensionality supersolids~\cite{Norcia2021}).

This paper is organized as follows: In Sec.~\ref{sec:2} we start with the mapping of the GPE onto a PFC type model. This is done for an arbitrary interaction in Sec.~\ref{sec:2A_equations}. Next, in Sec.~\ref{sec:2Aa_softcore}, we specify the interaction that is subsequently employed. Sec.~\ref{sec:2B_numerics} briefly explains the used numerical methods while Sec.~\ref{sec:2C_valid} discusses the validity of the approximated model for the nonlocal GPE. Sec.~\ref{sec:2E_patterns} presents bifurcation diagrams for one- and two-dimensional systems including branches of stable 
and unstable patterned and localized states. In Sec.~\ref{sec:3} we add a higher order nonlinearity to the model and subsequently discuss linear stability (Sec.~\ref{sec:3A}) as well as the resulting changes in the bifurcation diagram (Sec.~\ref{sec:3B}). Finally, in Sec.~\ref{sec:4}, we summarize and conclude. 

\section{\label{sec:2}Mapping the nonlocal GPE to a PFC model}
\subsection{\label{sec:2A_equations} Governing Equations and Mapping}

On the mean-field level the dynamics of a BEC is described by a nonlocal GPE for a complex scalar field $\Psi=\Psi(\mathbf{r},t)$, where the long-range interaction between the bosons gives rise to a convolution. On the mean-field level the nonlocal GPE can be written in rescaled form as~\cite{Gros1961nc,Pita1961spjetp}
\begin{equation}
i\partial_t\Psi(\textbf{r},t) = \left(-\frac{1}{2}\Delta+\int\text{d}\textbf{r}'|\Psi(\textbf{r}',t)|^2R(\textbf{r}-\textbf{r}')\right)\Psi(\textbf{r},t)\,. \label{eq:gp}
\end{equation}
Here, $R$ reflects both, long-range interatomic interactions and short-range interactions due to s-wave scattering.
Throughout this paper we use $\bar\rho=\frac{1}{V}\int|\Psi|^2\text{d}^nr$ as a free parameter. 
Our aim is to approximate Eq.~\eqref{eq:gp} in the vicinity of the phase transition in a way that does not involve a convolution. To approach this problem we start by considering the dispersion relation.

We employ the ansatz $\Psi(\textbf{r},t)=\left[\sqrt{\bar{\rho}}+\delta\psi(\textbf{r},t)\right]e^{-i\mu t}$ where $\sqrt{\bar{\rho}}>0$ is constant, $\mu=\bar{\rho}\int R(\textbf{r})\text{d}\textbf{r}$, and the perturbation with $|\delta\psi(\textbf{r},t)|\ll \bar{\rho}$ is $\delta\psi(\textbf{r},t)=ue^{i(\textbf{k}\textbf{r}-\omega t)}-v^*e^{-i(\textbf{k}\textbf{r}-\omega t)}$. After linearization of \eqref{eq:gp} in $\delta\psi$, the two coupled equations for $u$ and $v$ are solved for $\omega$ (see, e.g., Ref.~\onlinecite{KBRW2001pre,MMCP2013pra}). The resulting dispersion relation is
\begin{equation}
\omega^2({\bf k})=\frac{k^2}{2}\left(\frac{k^2}{2}+2\bar\rho \hat R({\bf k})\right)\,, \label{eq:disprel}
\end{equation}
where $\hat R({\textbf k})$ is the Fourier transform of $R({\textbf r})$. In the following we assume radial symmetry for simplicity and introduce $k=|\textbf{k}|$.

For a sufficiently large value of $\bar\rho$ and suitable $R({\textbf r})$ we expect the dispersion relation to show a classical ``maxon-roton structure'', i.e., starting from $\omega^2=0$ at $k=0$, $\omega^2$ first increases with $k$, then passes a maximum (the ``maxon''), and subsequently a minimum, called ``roton minimum'' (see, e.g., Fig.~\ref{fig:mapping}~(a)). The zero crossing of this minimum marks the onset of linear modulational instability of the uniform state. Above threshold there exists a band of unstable wavenumbers (i.e., with $\omega(k)^2<0$). Small periodic modulations on an otherwise plane background with $k$-values within this band grow exponentially.

The critical values $\bar\rho_c$ and $k_c$ at the threshold are given by $\omega^2(k_c)=0$ and $\dfrac{\partial \omega^2}{\partial k}\big|_{k=k_c}=0$. Our goal is to approximately match the given dispersion relation Eq.~\eqref{eq:disprel} at the stability threshold $(k_c,\bar\rho_c)$ using a Taylor series $\hat R^{(T)}(k)$:
\begin{equation}
 \hat R^{(T)}(k)=g_0+g_2 k^2+g_4 k^4 \label{eq:Rk}
\end{equation}
with the corresponding dispersion relation 
\begin{equation}
    \omega^{2(T)}=\frac{k^2}{2}\left(\frac{k^2}{2}+2\bar\rho \hat R^{(T)}\right).\label{eq:approx_disprel}
\end{equation} 
In analogy, one can introduce $k^{(T)}_c$ via $\omega^{{2(T)}}(k^{(T)}_c)=0$ and $\dfrac{\partial \omega^{2{(T)}}}{\partial k}\big|_{k=k^{(T)}_c}=0$.
To proceed we have to fix the three free parameters $(g_0,g_2,g_4)$ in an appropriate manner. 
As we want to approximate the behavior close to the linear stability threshold a reasonable request for the function $\omega^{{2(T)}}$ is that it should feature the same critical wavenumber $k_c$, the same critical density $\bar\rho_c$, and the same curvature in the roton minimum $\dfrac{\partial^2 \omega^{2}}{\partial k^2}\big|_{k=k_c}$. This can be realized by solving the set of equations given by
\begin{align}
 k_c&=k_c^{(T)}\,,\label{eq:req1}\\
 \bar\rho_c&=\bar\rho_c^{(T)}\,,\\
 \frac{\partial^2 \omega^{2}}{\partial k^2}\big|_{k=k_c} &=\frac{\partial^2 \omega^{2{(T)}}}{\partial k^2}\big|_{k=k^{(T)}_c}\,. \label{eq:req3}
\end{align}
Independently of the specifics of the long-range interaction, the formal solution of this set is given by 
\begin{align}
 g_0&=\frac{k_c^2\partial_{kk}\hat R(k_c)-2\hat R(k_c)}{8}\,,\nonumber\\
 g_2&=-\frac{2g_0-\hat R(k_c)}{k_c^2}=-\frac{k_c^2\partial_{kk}\hat R(k_c)-6\hat R(k_c)}{4k_c^2}\,,\nonumber\\
 g_4&=\frac{g_0}{k_c^4}\,.\label{eq:gs}
\end{align}
This differs from the approach in Ref.~\onlinecite{HeBD2019pra}, where $\hat{R}(k)$ is approximated based on its form up to the first minimum. These approaches are similar to gradient expansions of the linear term when deriving PFC models~\cite{ARRS2019pre} or to the weakly nonlocal limit in the context of optics.~\cite{KBRW2001pre}

The approximated version of Eq.~\eqref{eq:gp} now reads:
\begin{equation}
 i\partial_t\Psi = \left(-\frac{1}{2}\Delta+g_0|\Psi|^2-g_2\Delta |\Psi|^2+g_4\Delta^2|\Psi|^2 \right)\Psi\,. \label{eq:gp_approx}
\end{equation}
As the set $(g_0,\,g_2,\,g_4)$ is fixed by Eq.~\eqref{eq:gs}, this equation only depends on the same parameter $\bar{\rho}$ as before and $g_0,\,g_2,\,g_4$ are uniquely determined by the  geometry of the potential. 

Through a Madelung transformation,\cite{Made1927zp,CaDS2012n} $\Psi=\sqrt{\rho}e^{i\varphi}$ and separating real and imaginary parts we can rewrite Eq.~\eqref{eq:gp_approx} in the following fashion:
\begin{align}
\partial_t\varphi&= \left[\frac{\Delta\rho}{4\rho}-\frac{(\nabla\rho)^2}{8\rho^{2}}-\frac{(\nabla\varphi)^2}{2}-\left(g_0\rho-g_2\Delta\rho+g_4\Delta^2\rho\right)\right]\,,\label{equ:madelungphase}\\
\partial_t \rho&=-\nabla\cdot \left(\rho\nabla \varphi\right)\,.\label{equ:madelungrho}
\end{align}
An equation for steady states (including the ground states) is obtained employing the ansatz $\varphi=\mu t$, implying $\nabla\varphi=0$ and $\partial_t\varphi=\mu$. It reads
\begin{equation}\label{eq:steadyStates}
\mu=\left[\frac{\Delta\rho}{4\rho}-\frac{(\nabla\rho)^2}{8\rho^{2}}-\left(g_0\rho-g_2\Delta\rho+g_4\Delta^2\rho\right)\right]\,.
\end{equation}
To obtain the steady states dynamically via a numerical time simulation we follow appendix~D.1 of Ref.~\onlinecite{HeBD2019pra} and employ a dissipative, locally density-conserving dynamics, i.e., we use a continuity equation $\partial_t \rho=-\nabla\cdot {\bf j}$ as Eq.~\eqref{equ:madelungrho} but employing the flux $\mathbf{j}=\nabla \mu$. The resulting kinetic equation reads
\begin{align}
\partial_t \rho &= 
\Delta\left[\frac{1}{2}\left(\frac{(\nabla\rho)^2}{4\rho^{2}}-\frac{\Delta\rho}{2\rho}\right)+\left(g_0\rho-g_2\Delta\rho+g_4\Delta^2\rho\right)\right]\,.\label{equ:drhodt}
\end{align}
An alternative to the outlined approach to find groundstates  is called ``complex'' or ``imaginary'' time evolution (see e.g., Ref.~\onlinecite{ChST2000pre}), which employs a Wick-rotation $t\rightarrow-it$ and renormalizes the wavefunction after each propagation step. 

Note that the resulting dissipative Eq.~\eqref{equ:drhodt} corresponds to a PFC model \cite{ELWG2012aip,TARG2013pre} with an amended energy functional. Then, $\mu$ takes the role of a chemical potential. This observation allows one to relate the steady states and their bifurcations to corresponding results obtained for PFC models.

At this point we rescale the governing equation~\eqref{equ:drhodt} to obtain the conventional form for PFC-type equations -- apart from the peculiar form of the nonlinearity -- via 
\begin{align}
 r&\rightarrow r^\prime\,\sqrt{-\frac{g_2}{2g_4}}\,,\label{eq:rescaling1}\\
 t&\rightarrow t^\prime\,\frac{4g_4}{g_2^2}\,, \label{eq:rescaling2}\\
 \rho&\rightarrow \rho^\prime\,\frac{(-g_2)}{2}\,.\label{eq:rescaling3}
\end{align}
Using Eqs.~\eqref{eq:rescaling1}-\eqref{eq:rescaling3} and dropping the primes we find 
\begin{eqnarray}
    \partial_t \rho = \Delta\left[\frac{(\nabla\rho)^2}{8\rho^2}-\frac{\Delta\rho}{4\rho}+(\alpha_u+1)\rho+2\Delta\rho+\Delta^2\rho\right],\label{equ:main_eq}
\end{eqnarray}
where we use the same interaction parameter 
\begin{eqnarray}
 \alpha_u=\frac{4g_0g_4}{g_2^2}-1 \label{equ:alpha_u}
\end{eqnarray}
as in Ref.~\onlinecite{HeBD2019pra}. In the following, $\alpha_u$ will be a crucial parameter, so we add the convenient formula 
\begin{align}
    \alpha_u(k_c,\bar\rho_c)&=\frac{\left(k_c^2\partial_{kk}\hat{R}(k_c)-2\hat{R}(k_c)\right)^2}{\left(k_c^2\partial_{kk}\hat{R}(k_c)-6\hat{R}(k_c)\right)^2}-1\,.
\end{align}
It stems from Eqs.~\eqref{eq:gs} and directly links Eq.~\eqref{equ:main_eq} with the original Eq.~\eqref{eq:gp}.

For an appropriate mapping it is important to restrict the range of $\alpha_u$ to values that are in the interval $[-1,0]$. For smaller values, amplitudes become excessively large and lead to unphysical negative densities. For the interactions considered here, $\alpha_u<0$ follows naturally.

Written in gradient dynamics form, the dissipative conserved dynamics Eq.~\eqref{equ:main_eq} reads
\begin{equation}
    \frac{\partial\rho}{\partial t} = \Delta\frac{\delta F}{\delta \rho}\,,
\end{equation}
where the chemical potential is defined as the functional derivative 
\begin{equation}
  \mu = \frac{\delta F}{\delta \rho}
\end{equation}
of the functional
\begin{equation}
    F[\rho] = \frac{1}{2}\int_V  \left[\frac{|\nabla \rho|^2}{4\rho} +(\alpha_u+1)\rho^2+2\rho\Delta\rho+\rho\Delta^2\rho\right]\text{d}^nr\,. \label{equ:Ffunctional}
\end{equation}
Note, that in the following we adopt the vocabulary of soft-matter physics and call the functional a ``free energy functional'' despite the suppression of the temperature dependence of the original system. In our case, it stems from the fact that imposing $\nabla\varphi = 0$ for Eq.~\eqref{eq:steadyStates} implies that we consider dissipative pseudo dynamics. 

The corresponding grand potential density is $\Omega/V$ with
\begin{equation}
  \Omega =F - \int \mu\rho \text{d}^nr.
\end{equation}
Both $\mu$ and $\Omega/V$ are important for the identification of suitable parameter ranges where coexisting states can be found, which in turn indicate the occurrence of a first order phase transition.

\subsection{\label{sec:2Aa_softcore} Soft-core Bosons and their Dispersion Relation}
In the following, we consider a Heaviside step function $\Theta(1-r)$ with $r=|\textbf{r}|$ as prototypical toy model for soft-core interactions~\cite{HeNP2010prl,SaMB2012prl,MMCP2013pra} and a delta function $\delta(r)$ to describe s-wave scattering, hence 
\begin{equation}
    R(r)=\Theta(1-r)+a\delta(r)\,.
\end{equation}

Here, $a$ corresponds to the rescaled scattering length, and both the strength and range of the long-ranged interaction have been absorbed in the mean density.\cite{MMCP2013pra} Therefore, there is no loss of generality in assuming that the Heaviside step has range and amplitude of one.

We employ this particularly simple interaction to conveniently illustrate the method.  We do not discuss the subject of collapse that can occur for attractive scattering length, as it has been discussed elsewhere in depth.\cite{DCCR2001nature,MaSK2011nonlin} Here, we merely use the scattering length to deform the dispersion relation.

\begin{figure}
    \centering
    \includegraphics[width=0.99\columnwidth]{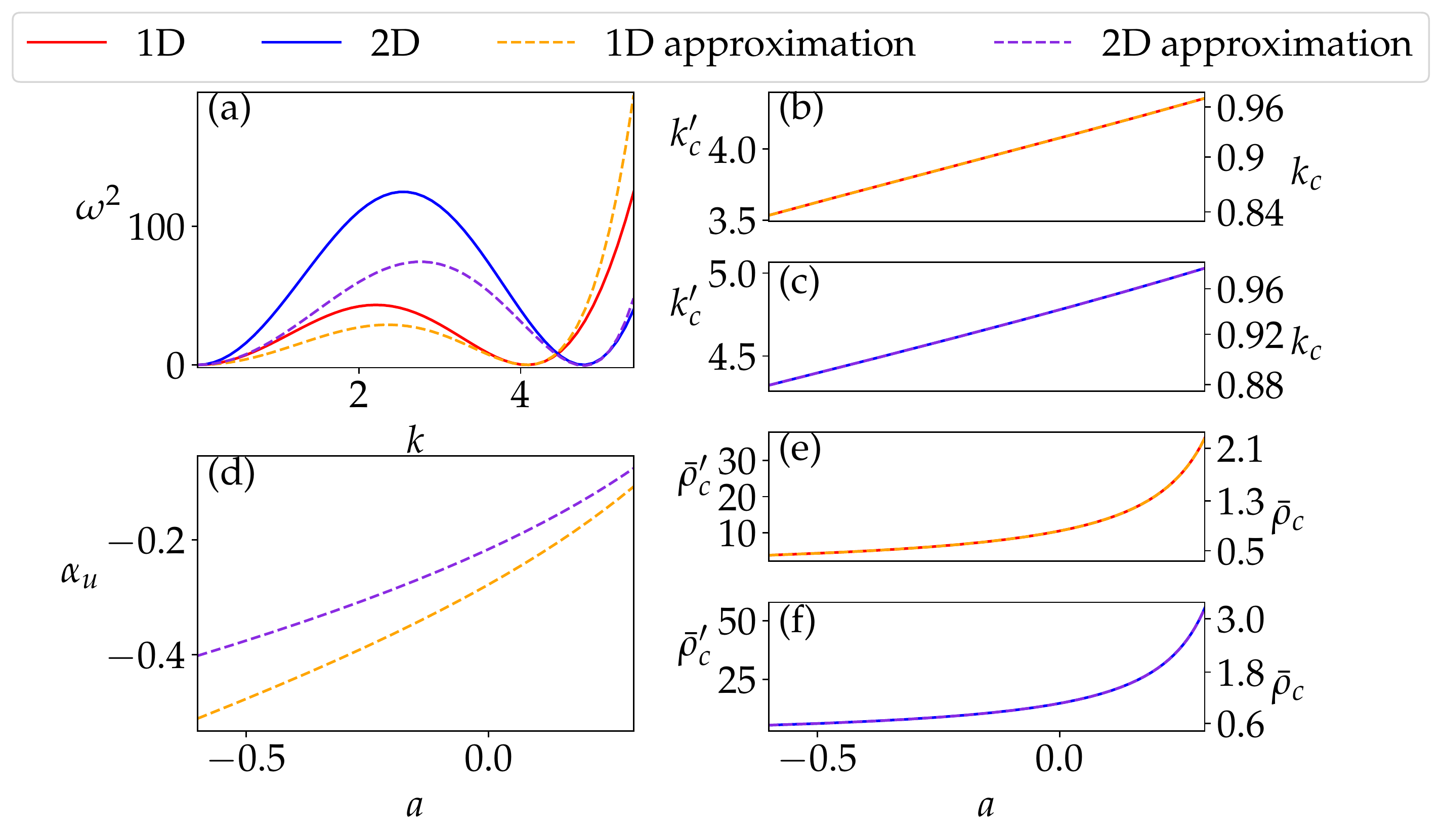}
    \caption{Comparison between the exact nonlocal model [Eq.~\eqref{eq:gp}] and its local approximation [Eq.~\eqref{eq:gp_approx} or Eq.~\eqref{equ:main_eq}]. (a) Dispersion relations given by Eqs.~\eqref{eq:disprel} (exact, solid lines) and \eqref{eq:approx_disprel} (approximation, dashed lines), both at the onset of the modulational instability, i.e.\ at $\bar\rho=\bar\rho_c$ and $\bar\rho^\prime=\bar\rho_c^\prime$, respectively. Furthermore, $a=0$. One- and  two-dimensional cases are represented in red/orange and blue/purple, respectively (see legend). Also given are the dependencies of the (b,c) critical wavenumber $k_c$ and the (e,f) critical density $\bar\rho_c$ on the rescaled scattering length $a$. Panel (d) shows the dependency of $\alpha_u$ on $a$. The two axis on the left and right of panels (b,c,e,f) relate to the unscaled and scaled system, respectively. As the scaling is nonlinear for both, $\bar{\rho}_c$ and $k_c$, the right axis is nonlinear.
    }
 \label{fig:mapping}
\end{figure}

The resulting dispersion relation Eq.~\eqref{eq:disprel} and its approximation Eq.~\eqref{eq:approx_disprel} are compared in Fig.~\ref{fig:mapping}. When evaluating Eqs.~\eqref{eq:gs} it has to be noted that they are implicit expressions: $k_c$ and $\bar\rho_c$ parametrically depend on $a$ and are obtained from Eq.~\eqref{eq:disprel}.
We find 
\begin{equation}
\hat{R}(k)=\begin{cases}
    \dfrac{2\sin(k)}{k} +a\,, & 1D  \\
    2\pi \dfrac{J_1(k)}{k} +a\,, & 2D 
\end{cases}
\end{equation}
where $J_1(k)$ is the Bessel function of the first kind. It is instructive to compare the dispersion relations of the approximated local model Eq.~\eqref{eq:approx_disprel} and the exact nonlocal model Eq.~\eqref{eq:disprel}. Figure~\ref{fig:mapping}~(a) compares the resulting dispersion relations at the onset of the modulational instability, i.e.\ at $\bar\rho=\bar\rho_c$ and $\bar\rho^\prime=\bar\rho_c^\prime$, respectively.
A specific value $a=0$ is chosen and one-dimensional (1D) and two-dimensional (2D) cases are both shown. Comparing the exact and approximate relations one sees that by construction, wavenumber $k_c$ and density $\bar{\rho}_c$ perfectly match while features like the curvature at $k=0$ and characteristics of the maximum deviate. Figures~\ref{fig:mapping}~(b) to~(f) furthermore show, how $k_c$, $\bar{\rho}_c$ and $\alpha_u$ depend on the rescaled scattering length $a$.
Before we proceed in the discussion of the validity of the approximated model in section~\ref{sec:2C_valid}, we comment on the  employed numerical methods.

\subsection{\label{sec:2B_numerics}Numerical Methods}
The ground state, i.e., the steady state with minimal energy of the GPE~\eqref{eq:gp} and its approximation [Eq.~\eqref{eq:gp_approx}]
can be obtained via direct numerical simulations using Fourier split-step method combined with the so-called complex time evolution. The latter consists in converting the original equation in a diffusion-type equation via a Wick rotation of the time coordinate, $t\rightarrow -it$ and renormalizing the norm after each step. In particular, the Fourier split-step method implies that the corresponding Hamiltonian is split, according to the Baker–Campbell–Hausdorff formula, into a linear kinetic part that is diagonal in Fourier space and a nonlinear part that is diagonal in real space. However, this method only gives access to stable ground states and does not permit to calculate any unstable solution branch. The robustness of the resulting solution can be tested by using Fourier split-step without the Wick rotation and renormalization (``real-time evolution'').

For the PFC type model with its dissipative pseudo dynamics (see Eq.~\ref{equ:main_eq}) it is possible to use a pseudospectral method in combination with semi-implicit Euler to find ground states.

While the results based on direct numerical integration allow one to obtain basic estimates of the stability range for a generic parameter set, a full bifurcation study of the system is still lacking. In order to track the pertinent states in parameter space we employ numerical pseudo-arclength continuation methods.~\cite{EGUW2019springer,KrOG2007springer} They trace branches of stable and unstable steady states through parameter space employing a combination of tangent predictor steps and correction steps using Newton's method. Here, we use path continuation employing the package \textsc{pde2path}.~\cite{UeWR2014cup}

If not stated otherwise, we employ periodic boundary conditions (BC) implying a valid continuous translational symmetry. Mean density conservation and phase condition (to exclude translational modes) are enforced via auxiliary conditions. The latter account for the continuous symmetries where the auxiliary conditions prevent a motion along the corresponding group orbit during continuation.~\cite{EGUW2019springer} To improve the performance of the path continuation in the 2D case, we choose domain geometry and size to match a specific pattern symmetry.
Occasionally, the periodic 2D domain can be further reduced by exploiting the symmetry of the pattern that is traced. For instance, a square pattern in a square domain may be considered in a domain one forth of the original size by employing Neumann BC that also automatically enforce the phase conditions for translation. However, bifurcations which break that symmetry are then no longer detected.

\subsection{\label{sec:2C_valid}Validity of the Approximated Model}
In the following, we compute the ground states and their bifurcations that mark transitions from uniform to modulated states for $R(r)=\Theta(1-r)$ (i.e., for $a=0$) for both, Eq.~\eqref{eq:gp} and Eqs.~\eqref{eq:gp_approx} [or Eq.~\eqref{equ:main_eq}] to probe the validity of the approximation. As path continuation now allows us to trace out unstable solutions of the approximated model, we will add them to the bifurcation diagrams as well. 

\begin{figure}
\includegraphics[width=\linewidth]{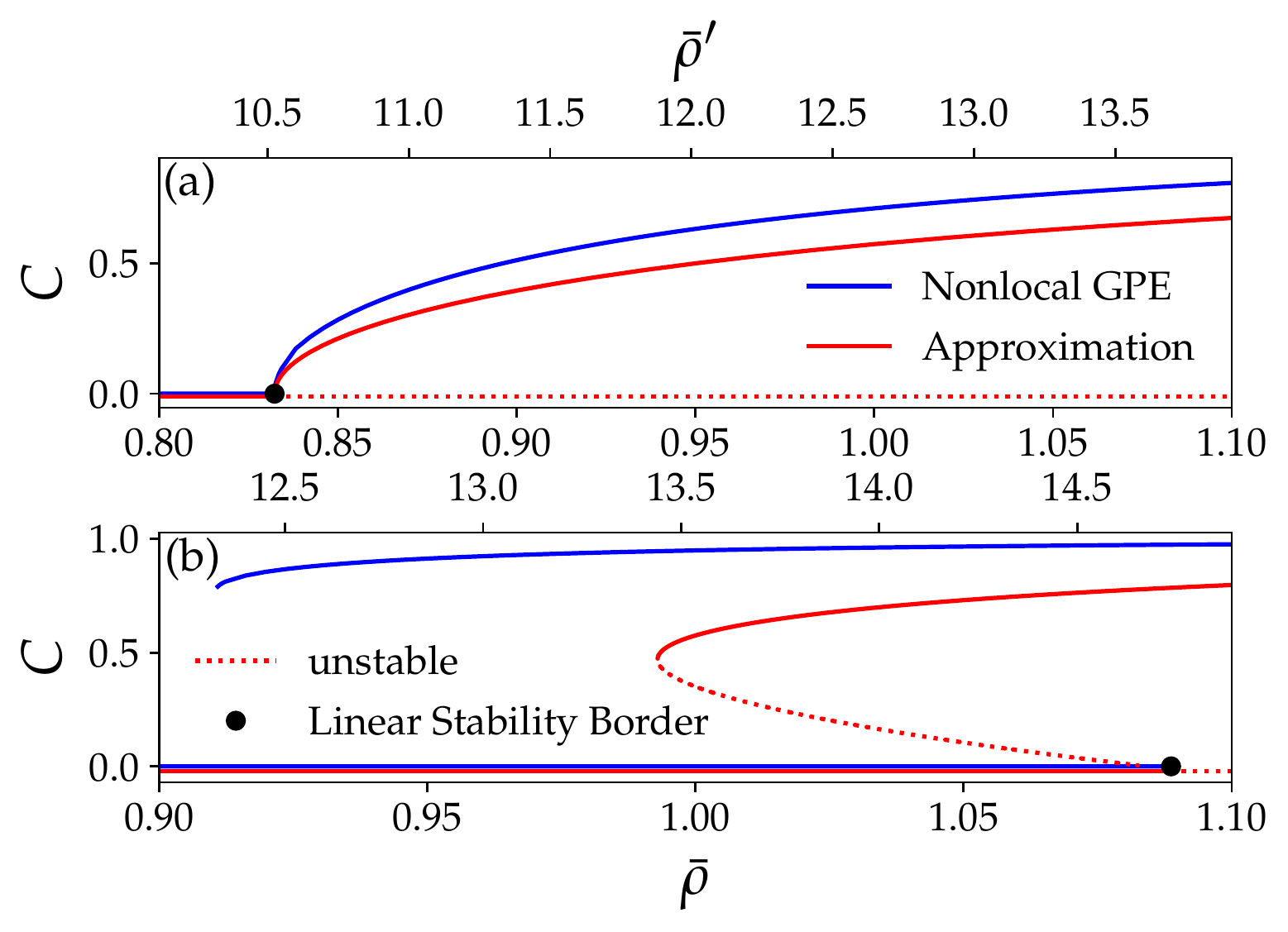} 
\caption{\label{fig:gp_vs_approx_1d} Comparison between steady states of the GPE [Eq.~\eqref{eq:gp}, blue line] and its approximation [Eq.~\eqref{equ:main_eq} or \eqref{eq:gp_approx}, red line]. The contrast $C$ [Eq.~\eqref{eq:contrast}] is shown for (a) 1D and (b) 2D systems 
as a function of the mean density $\bar\rho$ and $\bar\rho^\prime$. The domain size correspond to multiple of $2\pi/k_c$. The black filled circle indicates the linear stability threshold [cf.~Fig.~\ref{fig:mapping}~(b-d)].
}
\end{figure}

Figure~\ref{fig:gp_vs_approx_1d} shows the resulting bifurcation diagram for steady states in 1D and 2D in dependence of the mean concentration $\bar\rho$.%
\footnote{Note that here we always use mean density as control parameter when giving bifurcation diagrams. The corresponding Lagrange multiplier $\mu$ is adapted accordingly and could, in principle, be used as control parameter in an alternative bifurcation diagram. However, the proper control parameter ($\bar\rho$ or $\mu$) is determined by the dynamics or general set-up that one has in mind. If the dynamics conserves mass as in the here considered case, mean density is the correct control parameter to use as allowed perturbations do not change mass (but change chemical potential $\mu$ as it acts as Lagrange multiplier). Alternatively, if one uses the $\mu$ as control parameter one implicitly considers a nonconserved dynamics at fixed chemical potential. Then perturbations are at fixed chemical potential and mass changes. The bifurcation diagrams (and stabilities of otherwise identical states) are normally different in the two cases  (see e.g., discussions in the conclusion of Ref.~\onlinecite{TARG2013pre} and in section~2 of Ref.~\onlinecite{HAGK2021ijam}).}
The states are characterized by the contrast  
\begin{equation}
    C=\frac{\rho_\text{max}-\rho_\text{min}}{\rho_\text{max}+\rho_\text{min}}\label{eq:contrast}\,,
\end{equation}
where zero contrast represents a uniform superfluid state. As expected, the linear stability thresholds exactly agree, e.g., $\bar\rho_c'=10.5$ for the 1D case. Consistently between exact model and approximation, the primary bifurcation at $\bar\rho_c$, is supercritical in 1D and subcritical in 2D. In consequence, in 1D the weakly and fully nonlinear behaviors first closely agree and then quantitatively deviate further from the threshold once the amplitude of the density wave increases. The difference is about 20\% at the right hand limit of the diagram in Fig.~\ref{fig:gp_vs_approx_1d}~(a). In the 2D case, the linear stability threshold is at $\bar\rho_c'=14.7$ where the branch of patterned states subcritically emerges, leading to a region of bistability (not to be confused with thermodynamic coexistence which would indicate localized states). Note that the unstable states on the subcritical part of the branch can be obtained by path-continuation but not by complex time evolution. Here, we are only able to employ continuation for the approximate, i.e., local model, but not for the exact, i.e., nonlocal model. For the latter we 
obtain the uniform states analytically, while the stable patterned states, i.e., the states beyond the stabilizing saddle-node bifurcation, are numerically obtained by simulating the complex time evolution and probing robustness by real-time evolution (see Section~\ref{sec:2B_numerics}). Unfortunately, in this way the unstable subcritical part where we expect good agreement can not be compared. As the saddle-node bifurcation occurs in the fully nonlinear regime, there is an appreciable deviation (of about 10\% in absolute terms) between its positions in the exact and approximate model [see Fig.~\ref{fig:gp_vs_approx_1d}~(b)]. Note, that Ref.~\onlinecite{ARRS2019pre} also points out several closely related consequences of applying such an approximation to dynamic density functional theory, such as changes to phase boundaries and regions without physical solutions.

\section{\label{sec:2E_patterns}Periodic Patterns and Localized States}

In this section we will explore different types of periodic patterns as well as localized states supported by the approximated model~\eqref{eq:gp_approx} mainly in the 2D case. However, we will start with a brief discussion of the 1D case.

\subsubsection{\label{sec:2E_1d}Results for One Spatial Dimension}

In one spatial dimension and with a cubic nonlinearity in Eqs.~\eqref{eq:gp} and \eqref{eq:gp_approx} the bifurcation between a uniform perfect superfluid and a modulated state is typically supercritical (see Appendix~\ref{app:1mode}) as illustrated in the corresponding example in Fig.~\ref{fig:gp_vs_approx_1d}(a).

In case of a transition between states in a system without mass conservation the terms ``second order transition'' and ``supercritical bifurcation'' are closely related. However, there is no one-to-one relation in systems with mass conservation. There exist supercritical bifurcations in a bifurcation diagram with mean density as control parameter at parameter values that feature first order phase transitions as indicated by the coexistence of states.\footnote{Note that we use the notions of first and second order phase transition in the modern sense of discontinuous and continuous transitions, respectively (not in Ehrenfest's sense). We mainly use these notions when discussing the obtained phase diagrams. Then, first and second order refer to the existence and absence of a coexistence region, respectively.} This statement might seem contradictory, however, its validity rests in the fact that coexisting states normally represent different mean densities, i.e., are located at different values of the relevant control parameter. A direct consequence of such coexisting states is the possible presence of localized states. Such a presence can then be taken as indicative for the existence of a first order transition. To assess this possibility, one has to check  in detail whether states of identical chemical potential $\mu$ and grand potential $\Omega$ but different mean density exist. These are then the states coexisting in the thermodynamic limit. Note that regions of coexistence and regions of multistability, i.e., where different states are simultaneously linearly stable, do not coincide. A well analyzed example is provided by the phase field crystal model.\cite{TARG2013pre,TFEK2019njp,HAGK2021ijam}

\begin{figure}
\includegraphics[width=\linewidth]{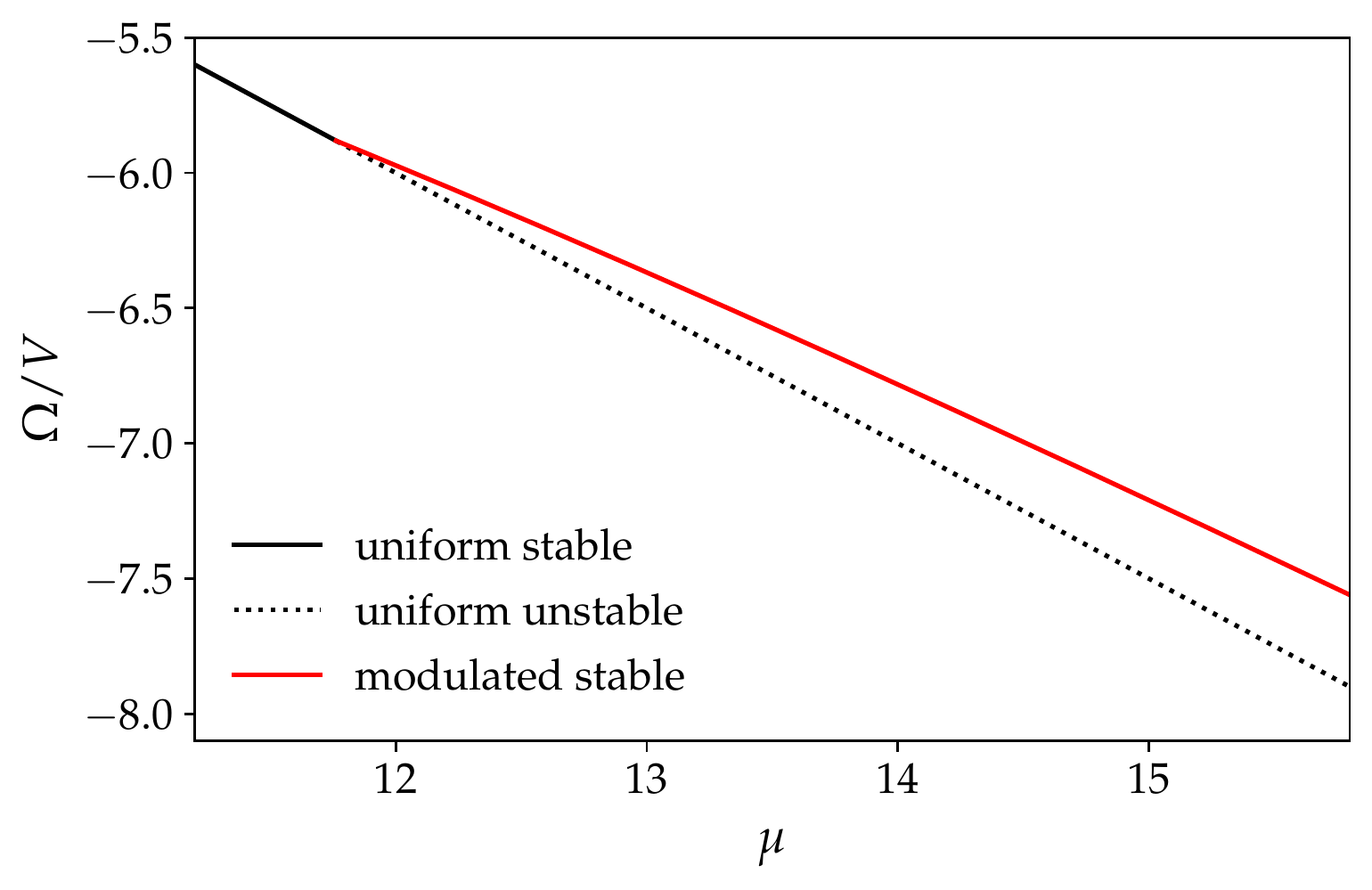}
\caption{\label{fig:1d_gp_mu-b} The grand potential density $\Omega/V$ is given as a function of the chemical potential $\mu$ for the patterned and uniform states in the approximated 1D case of Fig.~\ref{fig:gp_vs_approx_1d}~(a). Modulated and uniform states are given as red and black lines, respectively. Solid and dashed lines refer to stable and unstable states, respectively. As the lines do not cross, uniform and modulated states do not coexist and no localized states exist.}
\end{figure}

To test for coexistence, we plot in Fig.~\ref{fig:1d_gp_mu-b} the grand potential density $\Omega/V$ as a function of the chemical potential $\mu$ for the branches of uniform and patterned states presented in Fig.~\ref{fig:gp_vs_approx_1d}~(a) for the approximated model. In the representation of Fig.~\ref{fig:1d_gp_mu-b} the patterned and the uniform phase can coexist if their branches cross.~\cite{TFEK2019njp} As such a crossing does not occur, uniform and patterned states of different density will not coexist. This indicates that for the considered parameters the transition is indeed of second order and implies that no localized states can be expected in one spatial dimension.

\subsubsection{\label{sec:2E_2d}Results for Two Spatial Dimensions}
To determine patterns in two dimensions and their ranges of stability as a function of $\bar{\rho}$, we use pseudo-arclength continuation at different fixed values of $\alpha_u$. Thereby, the domain size is optimized for a hexagonal pattern matching the critical wavenumber $k_c$. The corresponding length is $L_c=\frac{2\pi}{k_c}$ resulting in the domain size $L_x\times L_y=\frac{2}{\sqrt{3}}L_c\times 2L_c$. Note that periodic BC are required, in particular, to accurately predict the stability of the stripe phase and related secondary bifurcations. Neumann BC do not allow for certain changes of symmetry, resulting in missing unstable modes and therefore bifurcations.\footnote{Bifurcations to intermediate states between stripes and hexagons directly influence the amount of positive eigenvalues found for the stripe state. Fig.~\ref{fig:2D_transition} shows that depending on the spatial phase of the stripes the pattern resulting from such a intermediate might not be mirror symmetric with respect to the $x$-axis. As the latter is required by Neumann BC the corresponding bifurcation is then missed. Therefore, Neumann BC ultimately give rise to the artefact that stripes of different spatial phase exhibit different stabilities.}

\begin{figure}
\includegraphics[width=\linewidth]{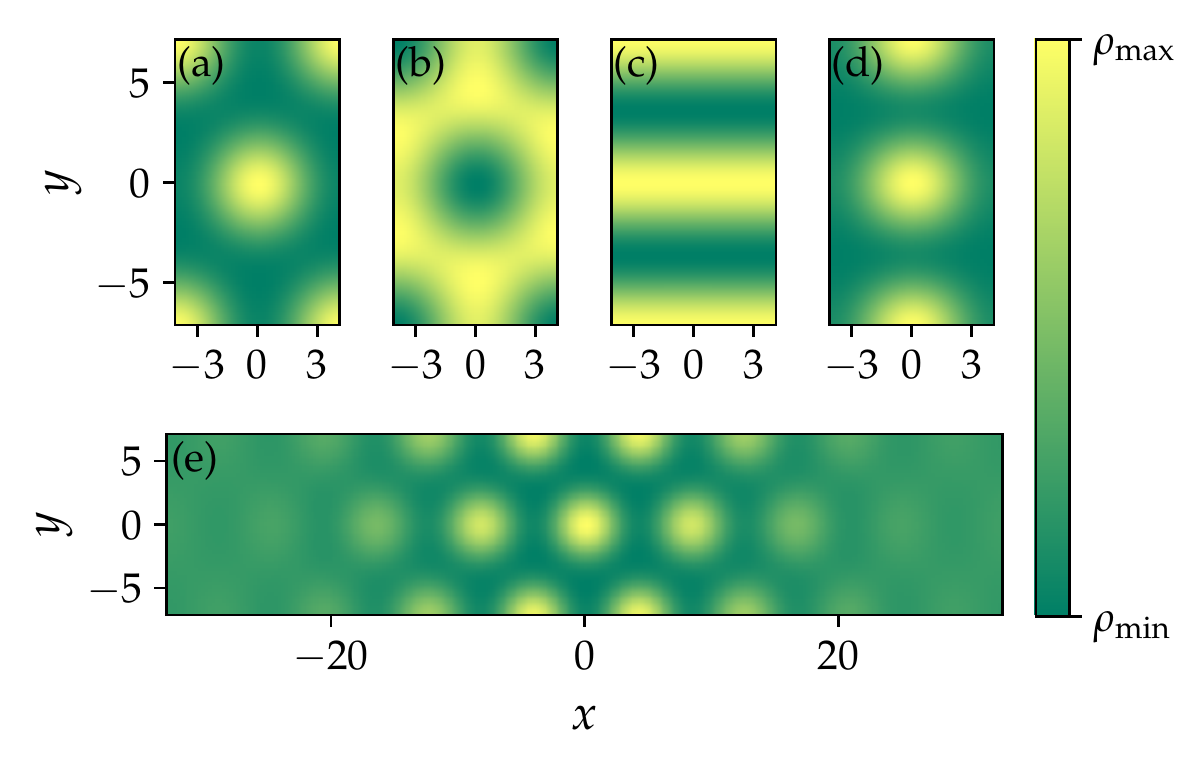} 
\caption{\label{fig:2D_patterns} Patterns occurring in a two-dimensional system at $\alpha_u=-0.4$: (a) hexagon, (b) honeycomb, (c) stripe (d) rectangle and (e) localized state corresponding to a patch of patterned state in  background of uniform ground state. The domain sizes $L_x\times L_y$ are  (a)-(d) $\frac{2}{\sqrt{3}}L_c\times 2L_c$ and (e) $16L_c/\sqrt{3}\times2L_c$.}
\end{figure}

Examples of the different occurring patterns are shown in Fig.~\ref{fig:2D_patterns}. In particular, we present up-hexagons (hexagonal pattern of peaks Fig.~\ref{fig:2D_patterns}~(a)), a honeycomb or down-hexagon pattern (hexagonal pattern of holes Fig.~\ref{fig:2D_patterns}~(b)), stripe pattern (Fig.~\ref{fig:2D_patterns}~(c)) and a rectangular pattern of peaks (Fig.~\ref{fig:2D_patterns}~(d)). The latter is a consequence of the rectangular domain allowing for hexagons. On a square domain, these would become square patterns, but no ideal hexagonal patterns would then exist.

\begin{figure}
\includegraphics[width=\linewidth]{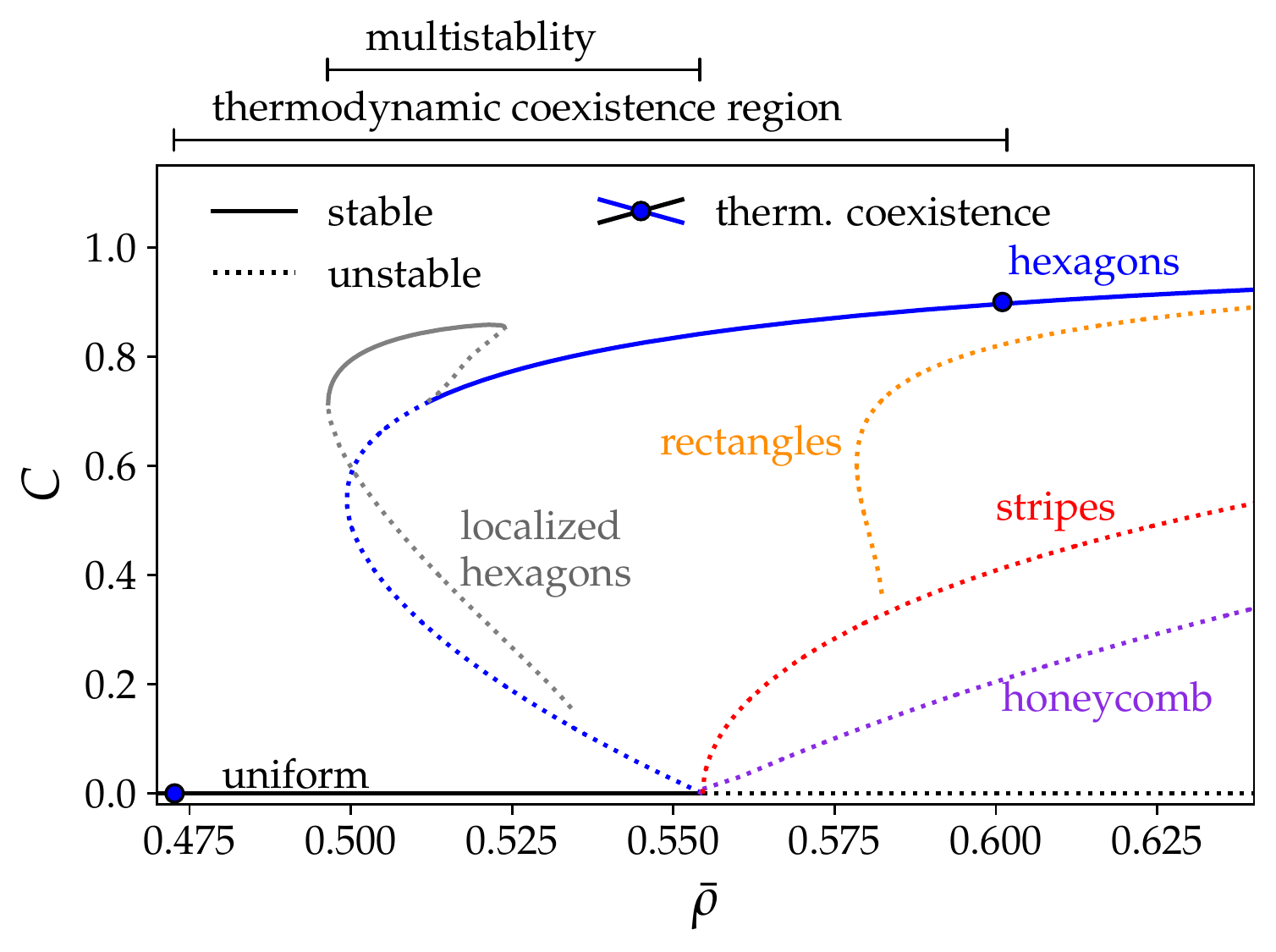} 
\caption{\label{fig:2d_bif} Bifurcation diagram of the steady state solutions in 2D of Eq.~\ref{equ:main_eq} as a function of $\bar{\rho}$ at fixed $\alpha_u=-0.4$. The domain sizes are as in Fig.~\ref{fig:2D_patterns}. The solutions are characterized through the contrast $C$ [see Eq.~\eqref{eq:contrast}]. Solid and dotted lines represent stable and unstable states, respectively. The respective patterns are indicated at the curves. The two filled blue circles represent stable states coexisting in an infinite domain (identical grand potential and chemical potential (see~Fig.~\ref{fig:2D_gp_mu}). This indicates the range where localized states (grey line) may, in principle, exist.}
\end{figure}

Figure~\ref{fig:2d_bif} presents the resulting bifurcation diagram giving $C$ as a function of the control parameter $\bar{\rho}$ as obtained at  $\alpha_u=-0.4$. The critical wave vector is $k_c=0.88$ leading to $L_c=7.14$. One can see that the uniform state undergoes a primary bifurcation at the expected critical value of $\bar{\rho}_c=0.555$. At this transcritical bifurcation, coming from low densities, the stable uniform state  loses stability and the unstable hexagon branch (blue line) at $\bar{\rho}<\bar{\rho}_c $ becomes the honeycomb branch (purple line) at $\bar{\rho}>\bar{\rho}_c $ that is also unstable. The hexagon branch passes a saddle-node bifurcation at $\bar{\rho}=0.498$ (where it stays unstable, but with fewer unstable modes) and turns back towards larger mean densities. It stabilizes at a pitchfork bifurcation at $\bar{\rho}=0.512$, i.e., below $\bar{\rho}_c$. This implies that there exists some hysteresis related to the multistability of different states. In practice, this can lead to a jump in the amplitude typical for a first order phase transition.

In addition to the hexagon/honeycomb branches an unstable branch of stripe patterns (red line) also emerges at the primary bifurcation. When ignoring all other branches the bifurcation would be classified as a supercritical pitchfork. 
In the considered finite rectangular domain, the branch of rectangular patterns (orange line) emerges in a subcritical pitchfork bifurcation from the branch of stripes. Both are always unstable. Note that in an infinite system it would correspond to a branch of square patterns and also emerge from in the primary bifurcation.

\begin{figure}
\includegraphics[width=\linewidth]{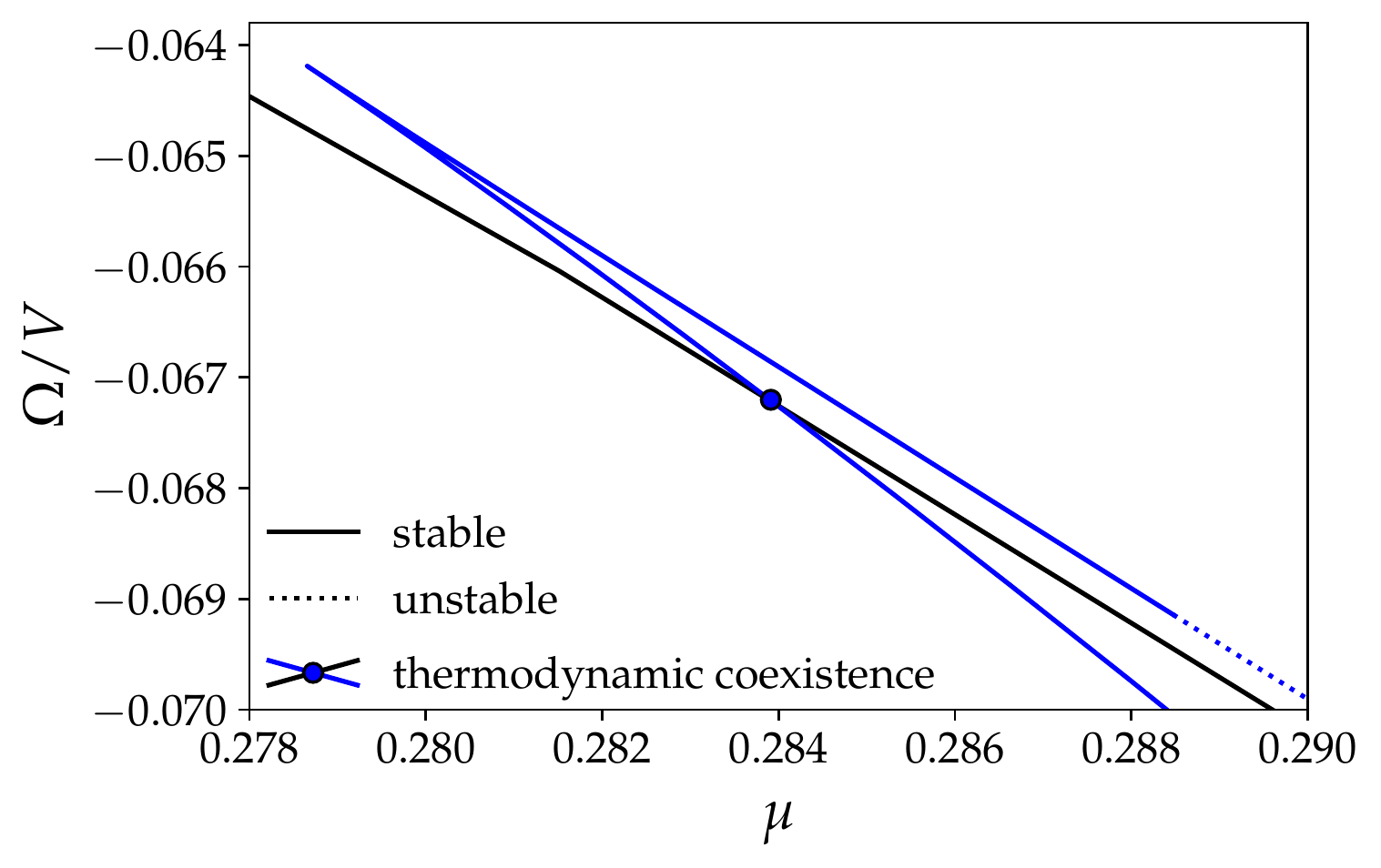} 
\caption{\label{fig:2D_gp_mu}  The grand potential density $\Omega/V$ as a function of the chemical potential $\mu$ for the branches of hexagonal patterns (blue lines) and uniform states (black lines) is shown for the 2D case in Fig.~\ref{fig:2d_bif}. The crossing of the lines marked by a filled circle indicates that these states can coexist and localized states may exist. The coexisting states are also marked in Fig.~\ref{fig:2d_bif}.}
\end{figure}

As in the 1D case we test the stable states for thermodynamic coexistence by plotting in Fig.~\ref{fig:2D_gp_mu} the grand potential density $\Omega/V$ as a function of the chemical potential $\mu$ for the corresponding branches. Indeed this time, in contrast to the 1D result, the corresponding lines cross indicating that stable uniform state and stable hexagonal pattern can coexist in an infinite domain. Note that the corresponding states have different densities and are marked in the bifurcation diagram Fig.~\ref{fig:2d_bif} by two filled blue circles. This indicates that at a sufficiently large domain size the system will exhibit localized states, for instance, for a domain size $16L_c/\sqrt{3}\times2L_c$  given by the gray line in Fig.~\ref{fig:2d_bif}. A corresponding profile is given in Fig.~\ref{fig:2D_patterns}~(e). Note that this localized state is only localized in $x$-direction, while it is periodic in $y$-direction. In case of a small 2D system which cannot host localized states, the hexagons stabilize directly at their saddle-node bifurcation, while the jump to the hexagons as the energetically favored state occurs at $\bar{\rho}=0.505$.

Coming back to the branch of localized states in Fig.~\ref{fig:2d_bif} we notice that it does not seem to exhibit the expected typical snaking behaviour found, e.g., in the PFC model.~\cite{TFEK2019njp} This is also the case for larger domain sizes of up to $24L_c/\sqrt{3}\times 2L_c$ where one would usually find pronounced snaking behaviour (see, e.g., Ref.~\onlinecite{HoAT2021jpcm}). However, Ref.~\onlinecite{TARG2013pre} shows that in the PFC model a larger effective temperature  eventually destroys the pronounced snaking that involves saddle-node bifurcations of the branches of localized states. The number of peaks still changes when going along the branch, however, all saddle-node bifurcations have been annihilated. We believe that in the present case the value of $\alpha_u$ is too high for snaking to be found. Indeed, lowering $\alpha_u$ leads to steeper transitions between the uniform background and the patterned part within the localized states.  However, as the system reacts very  sensitively to a decrease in $\alpha_u$, we are not able to lower it significantly further due to the occurrence of negative densities that result in a break down of the numerical procedure. Finally, the lack of a second branch of localized states with an even number of lines of peaks (cf.~Fig.~15~(d) of Ref.~\onlinecite{TFEK2019njp})
is due to the numerical limitation of the size of the box and the choice of Neumann BC [Fig.~\ref{fig:2D_patterns}~(e)]. 

\begin{figure}
\includegraphics[width=\linewidth]{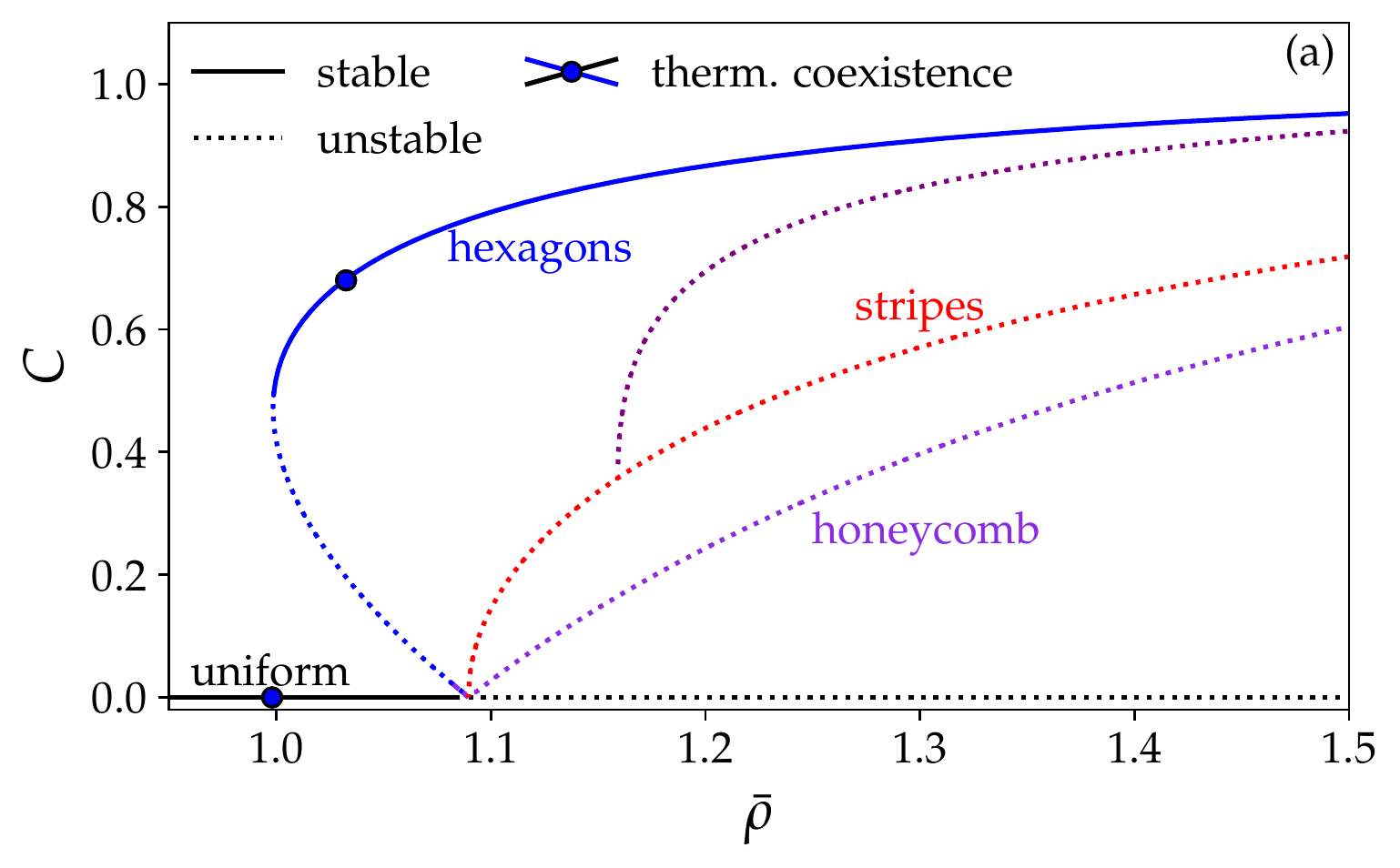}\\
\includegraphics[width=\linewidth]{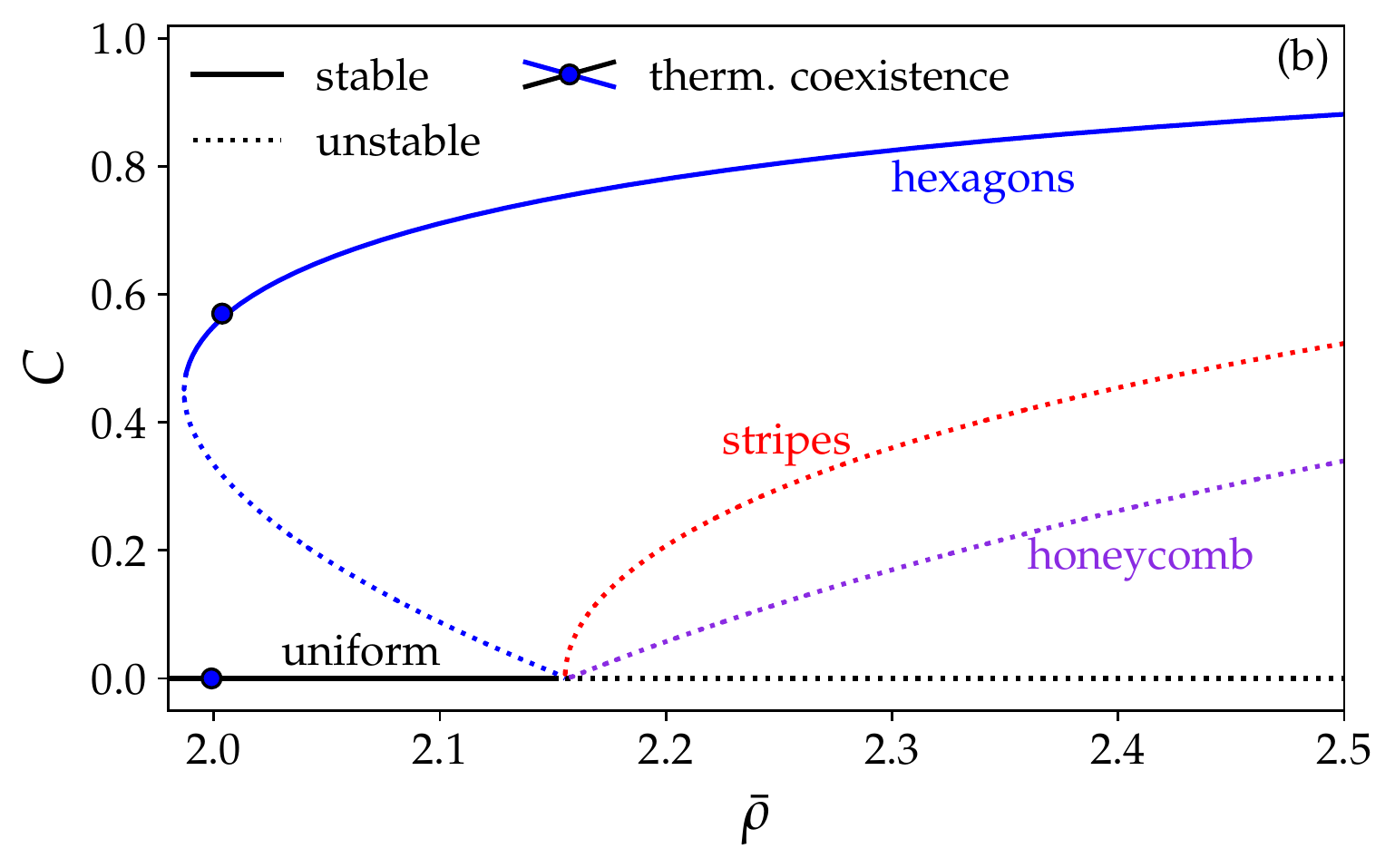} 
\caption{\label{fig:2d_bif_au022} 
Bifurcation diagrams of steady states of Eq.~\ref{equ:main_eq} in 2D as a function of $\bar{\rho}$ at fixed (a) $\alpha_u=-0.216$ and (b) $\alpha_u=-0.113$. The domain size is $2L_c/\sqrt{3}\times2L_c$. 
Solution measure, line styles, markers and pattern types are as in Fig.~\ref{fig:2d_bif}. The purple line in (a) corresponds to a superlattice-like pattern, namely, a rectangular arrangement of peaks with these rectangles being located on a hexagonal lattice}
\end{figure}

Further increasing the value of the interaction parameter to $\alpha_u=-0.216$ and $\alpha_u=-0.113$ results in the bifurcation diagrams shown in Figs.~\ref{fig:2d_bif_au022}~(a) and \ref{fig:2d_bif_au022}~(b), respectively.
First, inspecting Fig.~\ref{fig:2d_bif_au022}~(a) and comparing it to Fig.~\ref{fig:2d_bif} one notes that the branch of localized states does not exist anymore. In consequence, the hexagonal patterns become linearly stable directly at the saddle-node bifurcation. The bifurcation from the stripe state to the rectangles is replaced by a supercritical pitchfork bifurcation resulting in states that transform the stripes into an unstable superlattice-like pattern, namely, a rectangular arrangements of peaks with these rectangles being situated on a hexagonal lattice. This is allowed for by the Neumann BC applied to the same domain size as before used with the periodic BC. This effectively corresponds to a doubled domain size with periodic BC.~\cite{CGGK1991lnm} The second length scale present in the superlattice pattern is not intrinsic to the model, but reflects the domain size. In other words, the confinement creates defects in the pattern that are then repeated periodically due to the boundary condition. The bifurcation's relative distance to the primary bifurcation is slightly larger than at $\alpha_u=-0.4$. Second, focusing on Fig.~\ref{fig:2d_bif_au022}~(b), the picture has simplified more as an increase in $\alpha_u$ rapidly moves the secondary bifurcation on the branch of stripes towards large densities outside the $\bar\rho$-range of Fig.~\ref{fig:2d_bif_au022}.

In both cases, the primary bifurcation occurs at the expected value, namely, 
at $\bar{\rho}=1.09$ and $\bar{\rho}=2.15$ in Fig.~\ref{fig:2d_bif_au022}~(a) and (b), respectively. The saddle-node bifurcation on the branch of hexagonal patterns are at $\bar{\rho}=1.99$ and $\bar{\rho}=0.99$, respectively, i.e., with increasing $\alpha_u$ it moves further away from the primary bifurcation (in terms of relative density).

Summarizing, one may say that the range of bistability of hexagons and the uniform state increases with increasing interaction parameter $\alpha_u$. However, in contrast to this the actual range of thermodynamic coexistence between the two states shrinks with increasing $\alpha_u$ as indicated by the decreasing density difference between the respective two filled circles in Figs.~\ref{fig:2d_bif}, \ref{fig:2d_bif_au022}~(a) and \ref{fig:2d_bif_au022}~(b) that indicate the coexisting states, i.e., the positions of the binodals. Also note that for the two larger $\alpha_u$ both respective binodal densities are in the density range where the uniform state is linearly stable. Thermodynamically speaking, the spinodal  does not lie between the binodals. This is in contrast to the PFC results of Refs.~\onlinecite{TARG2013pre,TFEK2019njp} and resembles the behaviour encountered in density functional theories of colloidal crystallization (cf.~e.g.~Fig.~2 of Ref.~\onlinecite{AWTK2014pre}).

Even for a narrow coexistence region one would still expect localized states to exist. We believe that they could only be found for the parameters of Fig.~\ref{fig:2d_bif_au022}(b) if a much larger domain size was used as at present possible for us. In accordance with Ref.~\onlinecite{TARG2013pre} the increase in $\alpha_u$ leads to wider fronts between the pattern and the uniform state implying the need for larger domains.

However, to substantiate our hypothesis that the existence of localized states in the PFC approximation is indicative for their existence in the nonlocal GPE we have performed complex time evolution of Eq.~\eqref{eq:gp} as detailed in Appendix~\ref{app:loc}. In this way we could identify parameter sets for which localized states indeed represent the ground state of the nonlocal system.

\section{\label{sec:3}Higher-Order Nonlinearities}

In the context of dipolar BECs it has been shown that deep quantum effects, so-called quantum fluctuations,~\cite{LiPe2011pra} can lead to qualitative changes in the stability properties of the accessible patterns.~\cite{ZhMP2019prl,ZhPM2021pra,HSGB2021prr} Furthermore, they can give rise to qualitative changes in the character of the phase transition: The hexagonal crystallisation into a supersolid can be of second order \cite{ZhMP2019prl} in contrast to the first order transition described above. 
Here, we qualitatively explore features facilitated by a phenomenological addition of a higher-order nonlinearity to the equation of motion. This shall allow for an improved understanding of the emergence of the different phases and of accompanying changes in the bifurcation diagram. Note, however, that for illustration purposes the parameter values we consider are significantly larger than the values one would obtain in an actual computation of beyond mean-field corrections for the specific interaction considered here.

The considered nonlinearity has the form $|\Psi|^3\Psi$ with prefactor $b>0$. Our (rescaled) governing equation \eqref{eq:gp_approx} of motion is modified to
\begin{equation}
 i\partial_t\Psi= \left(-\frac{\Delta}{2}+(\alpha_u+1)|\Psi|^2+2\Delta|\Psi|^2
 +\Delta^2|\Psi|^2 + b|\Psi|^3\right)\Psi.\label{equ:beyond_mean_psi}
\end{equation}
The free energy functional defined by Eq.~\eqref{equ:Ffunctional} as well as the steady state equation~\eqref{equ:drhodt} are modified accordingly. We have
\begin{align}
& F = \int_V \text{d}^nr \left[\frac{|\nabla \rho|^2}{8\rho} + b\frac{2}{5}\rho^{5/2} +\frac{(\alpha_u+1)}{2}\rho^2+\rho\Delta\rho+\frac{\rho}{2}\Delta^2\rho\right] \label{equ:Ffunctional_b}
\end{align}
and
\begin{align}
&\partial_t \rho =
	\Delta\left[\frac{(\nabla\rho)^2}{8\rho^2}-\frac{\Delta\rho}{4\rho}+(\alpha_u+1)\rho+2\Delta\rho+\Delta^2\rho+b
	\rho^{3/2}\right]\,. \label{equ:beyond_mean}
\end{align}
In consequence, the equations~\eqref{eq:gs} are no longer suitable and have to be adapted accordingly. This is done in Appendix~\ref{app:g2j+qmf}.
Next, we investigate the changes caused by the unusual scaling of the additional nonlinearity.
\subsection{\label{sec:3A}Influence on Linear Stability}
\begin{figure}
\includegraphics[width=\linewidth]{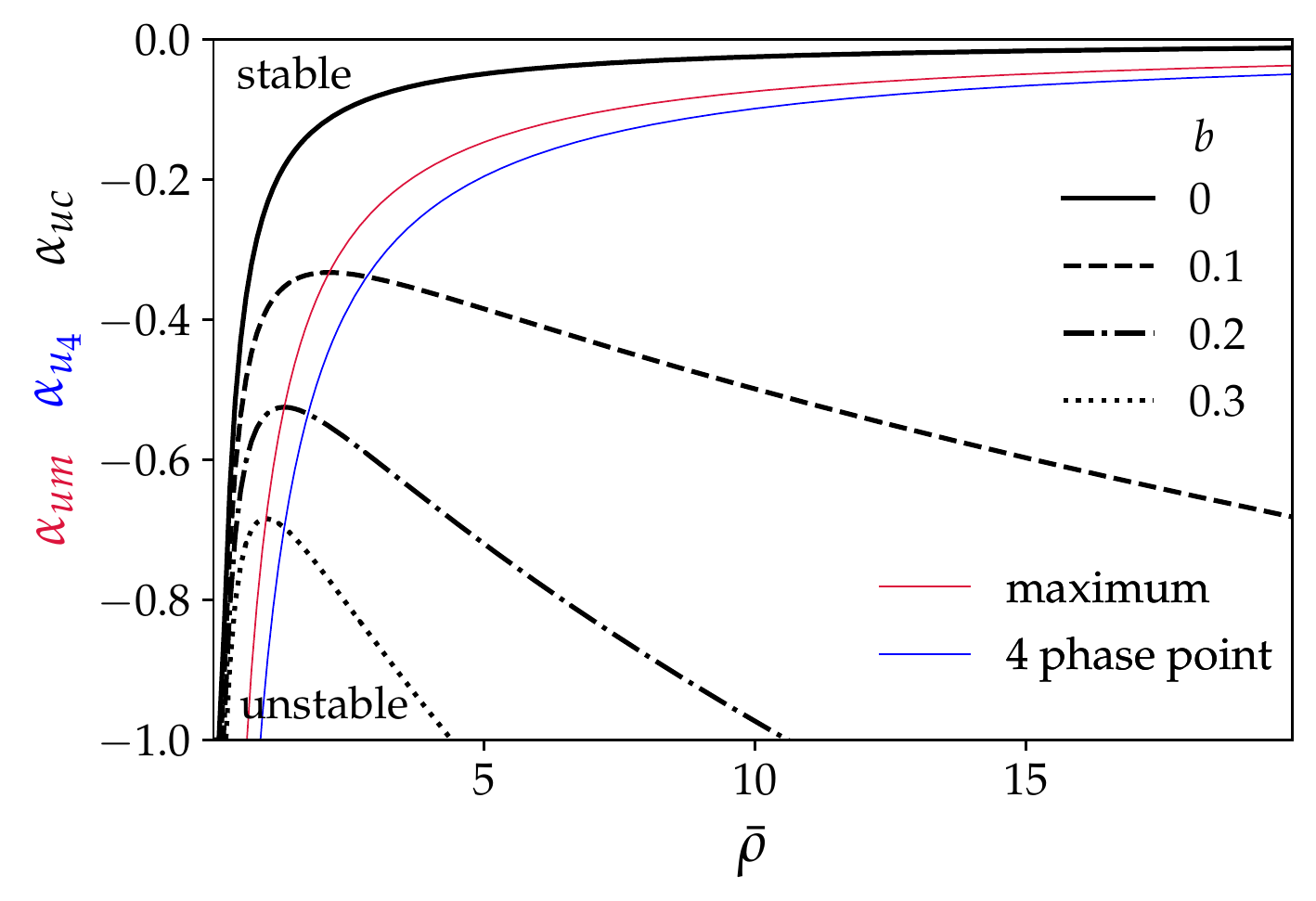} 
\caption{\label{fig:lin_stab_b} Linear stability region of the uniform (superfluid) state $\rho(x)=\bar{\rho}$ in the $(\bar{\rho},\,\alpha_u)$-plane. The stability thresholds according to Eq.~\eqref{eq:stab_bord_b} are given as (solid, dashed, dot-dashed and dotted) black lines for various values of the strength of the quantum fluctuation term $b$ as given in the legend. The regions above the respective lines correspond to stable uniform states. The thin red line indicates the position of the maximum $\alpha_{um}$ of the curves when continuously changing $b$. Note that for $b\to0$ the corresponding $\bar{\rho}$ also diverges. Correspondingly, the thin blue line indicates the position of the four-phase point $\alpha_{u4}$ where the phase transition is of second order (for details see main text). To the left [right] of the blue line the first stable pattern is the hexagon [honeycomb]. In both cases the transition is of first order.}
\end{figure}

Following the steps explained in Section~\ref{sec:2A_equations} we obtain the amended dispersion relation 
\begin{eqnarray}
\omega^{\prime2}(k)=\omega^2+\frac{k^2}{2}3b\bar\rho\sqrt{\bar\rho}\,.\label{eq:disp_beyond}
\end{eqnarray}
Perturbations grow exponentially when $\omega^{\prime2}<0$ with the threshold of instability at $\omega^{\prime2}=0$ which coincides with the emergence of nontrivial roots other than $k_0=0$:
\begin{eqnarray}
k_0^2 = 1-\frac{1}{8\bar{\rho}}\pm\sqrt{\left(1-\frac{1}{8\bar{\rho}}\right)^2-\left(\alpha_u+1+\frac{3}{2}b\sqrt{\bar{\rho}}\right)}\,.\label{eq:k0_b}
\end{eqnarray}
Therefore in a system of infinite size the threshold of instability, where only one $k_0=k_c$ exists, can be described through 
\begin{eqnarray}
 \alpha_{uc}=\left(1-\frac{1}{8\bar{\rho}}\right)^2-1-\frac{3}{2}b\sqrt{\bar{\rho}} \qquad &\text{for} \qquad\bar{\rho}\geq\frac{1}{8}\,.
 \label{eq:stab_bord_b}
\end{eqnarray}
where $\alpha_{uc}$ is the critical value of $\alpha_u$. It is depicted in Fig.~\ref{fig:lin_stab_b}.

As expected from the arguments laid out in Ref.~\onlinecite{ZhMP2019prl}, the higher-order (quartic) nonlinearity in Eq.~\eqref{equ:beyond_mean_psi}  significantly changes the phase diagram. This is illustrated in Fig.~\ref{fig:lin_stab_b} which shows the critical $\alpha_{uc}$ as a function of $\bar\rho$. At $b=0$ we see that $\alpha_{uc}(\bar\rho)$ has no local maximum, instead the maximum is at infinite $\bar\rho$. 
However, for nonzero values of $b$ the maximum at $\alpha_{um}$ moves to finite values of $\bar\rho$ -- its changing position is indicated by the red line. In consequence, for increasing $b$ the region of stable uniform states is extended towards smaller values of $\alpha_{uc}$ and larger values of $\bar{\rho}$. The blue curve indicates the single point at each value of $b$ where the phase transition is of second order. As explained below, this is computed from a one-mode approximation. In other words, the line marks the parameter values where the transcritical bifurcation between uniform and hexagon/honeycomb pattern corresponds to orthogonally crossing branches. At this point there is no multistability or coexistence of states. In consequence, to the left of the blue line the first stable pattern is a hexagon (or possibly a localized hexagon state) while to the right it is a honeycomb (possibly localized).

\subsection{\label{sec:3B}Changes to Patterns and Phase Transition}

To confirm the expected changes~\cite{ZhMP2019prl,ZhPM2021pra,HSGB2021prr} caused by the additional higher-order nonlinearity in existence region and stability of different patterns and the corresponding coexistence and phase transitions we apply a one-mode approximation (for details see Appendix~\ref{app:1mode}). In other words, the patterned state is approximated by harmonics of a single wavelength with small amplitude $A$. Inserting this ansatz into the free energy $F$ combined with a Taylor expansion in $A\ll1$ gives
\begin{eqnarray}
 F-F_0 = q_2A^2+q_3A^3+q_4A^4\label{eq:1mode}
\end{eqnarray}
which is a reasonable approximation close to the border of linear stability where the modulation amplitude $A$ is small. Minimizing $F$ with respect to $k$ gives the critical wave number $k_c$ and allows one to obtain the expressions for $q_2$, $q_3$ and $q_4$ given in Table~\ref{tab:one-mode_FA} in Appendix~\ref{app:1mode}. Minimizing $F$ with respect to $A$ one then determines the state of lowest energy.~\cite{ZhMP2019prl}

\begin{figure}
\includegraphics[width=\linewidth]{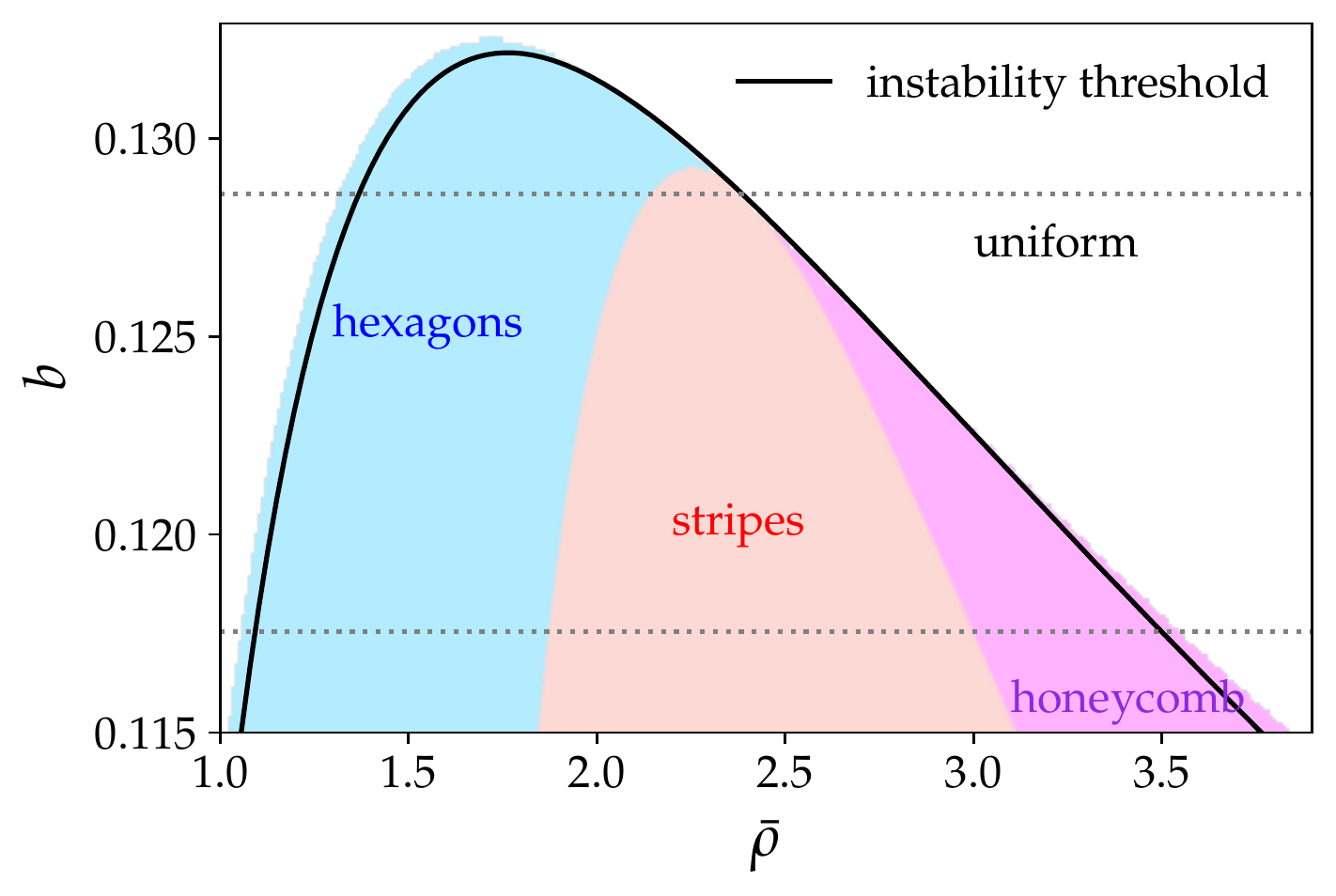} 
\caption{\label{fig:pt_change} Energetically favored ground states in the $(\bar\rho,b)$-plane as obtained in a one-mode approximation at $\alpha_u=-0.4$. The border of stability, Eq.~\eqref{eq:stab_bord_b}, is given as black line. The ground states are the uniform state (white region), hexagons (light blue), stripes (light red) and honeycombs (light purple). All four meet in a single point, where no bistability with the uniform state exists. Horizontal dashed lines indicate $b$-values employed for the bifurcation diagrams in Figs.~\ref{fig:2D_bif_b} and \ref{fig:2D_bif_b_hon}.}
\end{figure}

Alternatively, Fig.~\ref{fig:pt_change} gives an overview of the dominant phases in the $(\bar\rho,b)$-plane also obtained in the one-mode approximation. As we only consider a polynomial in $A$ up to fourth order it only provide an approximate phase diagram. The hexagonal pattern forms the ground state in the low-density part of the region of linearly unstable uniform states. In contrast, in the high-density part the honeycomb pattern has lowest energy while at intermediate densities the stripe pattern dominates. Note that both, hexagon and honeycomb regions of dominance extend beyond the border of linear stability of the uniform state. This suggests regions of bistability and coexistence indicating a first order phase transition. The bistability region shrinks when approaching the ``four-phase point'' where the borders of thermodynamic stability meet for all four stable states. When crossing the phase boundary at this point there is no bistability or coexistence indicating a second order phase transition.

The four-phase point is determined as an intersection of $q_3=(16b\bar{\rho}-8+1/\bar{\rho})3/256=0$ and Eq.~\eqref{eq:stab_bord_b}. The latter relation characterizes the primary bifurcation and the former states that the energy $F$ only depends on even powers of the pattern amplitude, so neither hexagons (positive $A$) nor honeycombs (negative $A$) are favored. We find its location $(\alpha_{u4},\bar\rho_4)$ as function of $b$ or alternately $(b_4,\bar\rho_4)$ as function of $\alpha_{u}$ by solving
\begin{eqnarray}
         b_4 &=\dfrac{8\bar{\rho}-1}{16\bar{\rho}^{5/2}}\\
 \alpha_{u4} &= \dfrac{7}{64 \bar{\rho}^2}-\dfrac{1}{\bar{\rho}}
\end{eqnarray}
The blue line in Fig.~\ref{fig:lin_stab_b} marks the resulting location of the four-phase point in the $(\bar\rho,\alpha_{u})$-plane for continuously changing $b$. Note that the four-phase point actually corresponds to a thermodynamic critical point. Its location within the phase diagram in Fig.~\ref{fig:pt_change} reflects that the parameters $b$, $\alpha_u$ and $\bar\rho$ are not independent.

\begin{figure}
\includegraphics[width=\linewidth]{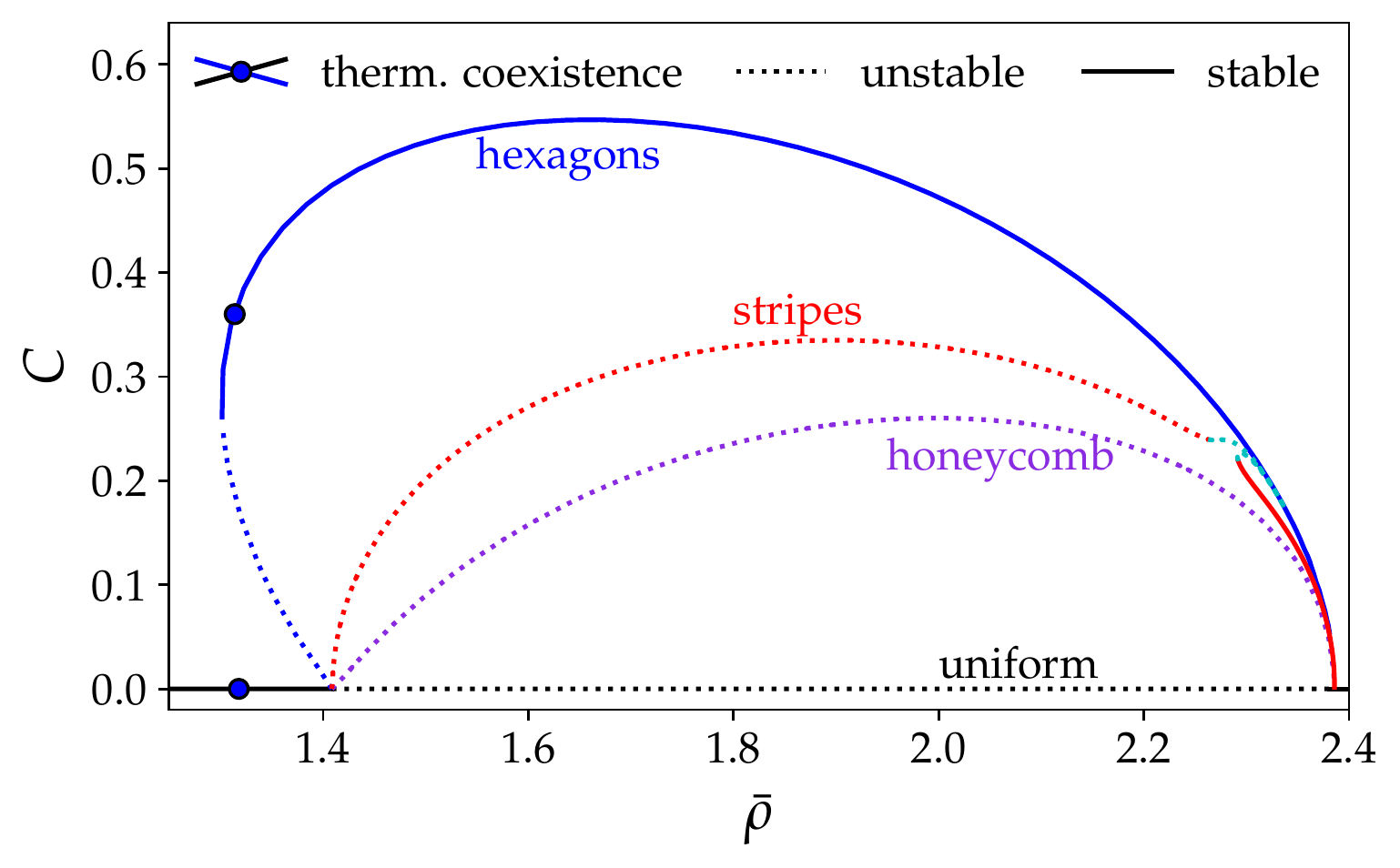} 
\caption{\label{fig:2D_bif_b} Bifurcation diagram showing branches of steady states of Eq.~\eqref{equ:beyond_mean} as a function of $\bar{\rho}$ at fixed $\alpha_u=-0.4$ and $b=0.1286$ (along the upper horizontal line in Fig.~\ref{fig:pt_change}). The considered 2D system has domain size $2L_c/\sqrt(3)\times2L_c$. The solutions are characterized through the contrast $C$ (see Eq.~\eqref{eq:contrast}). Line styles are as in Fig.~\ref{fig:2d_bif_au022} with the addition intermediate states between stripes and hexagons in cyan. The bifurcation at low $\bar\rho$ is related to points of thermodynamic coexistence between the uniform and hexagon states (indicated by filled circles). No such coexistence occurs related to the bifurcation at high $\bar\rho$ indicating a second order transition.}
\end{figure}

The predicted existence of different transitions is highlighted in the corresponding bifurcation diagram with $\bar{\rho}$ as control parameter as given in Fig.~\ref{fig:2D_bif_b}. It shows the branches of the emerging states of a two-dimensional system with $\alpha_u=-0.4$ at $b=0.1286$ (along the upper horizontal line in Fig.~\ref{fig:pt_change}). The parameters are chosen in such a way that at low $\bar\rho\approx1.35$ both, a bistability region and a (smaller) coexistence region between uniform state and hexagons are found indicating a first order phase transition. In contrast, at higher $\bar\rho\approx2.4$ the four-phase point is passed, i.e., at this specific value of $b$, the corresponding bifurcation is of higher codimension and stable hexagons and stripes both emerge vertically. The absence of a coexistence region indicates a second order transition. The domain size is again chosen to fit hexagonal states with $2L_c/\sqrt{3}\times2L_c$.

At lower densities we see the expected transcritical bifurcation. Again, the filled circles indicate coexisting uniform and hexagon states, i.e., states of identical chemical potential and grand potential density. This indicates a first order transition and implies that localized states should in principle exist. However, they are not observed as the domain sizes we are able to consider are not sufficiently large.
At higher densities, however, the bifurcation occurs on the critical point, i.e., at the specific parameter value where the transcritical bifurcation between hexagons and honeycombs and the related saddle-node bifurcation of the hexagons exactly coincide. As there are no points of identical chemical potential and grand potential density, there are no points of coexistence of these states with the uniform state beside the bifurcation point itself, and the corresponding phase transition is of second order. Note furthermore that the honeycomb state is unstable everywhere on the branch, while the stripe state has a narrow linearly stable region at high densities. However, in contrast to the result of the one-mode approximation (Fig.~\ref{fig:pt_change}), it never becomes the energetically favored state, but only reaches the same low energy as the hexagons.

\begin{figure}
\includegraphics[width=\linewidth]{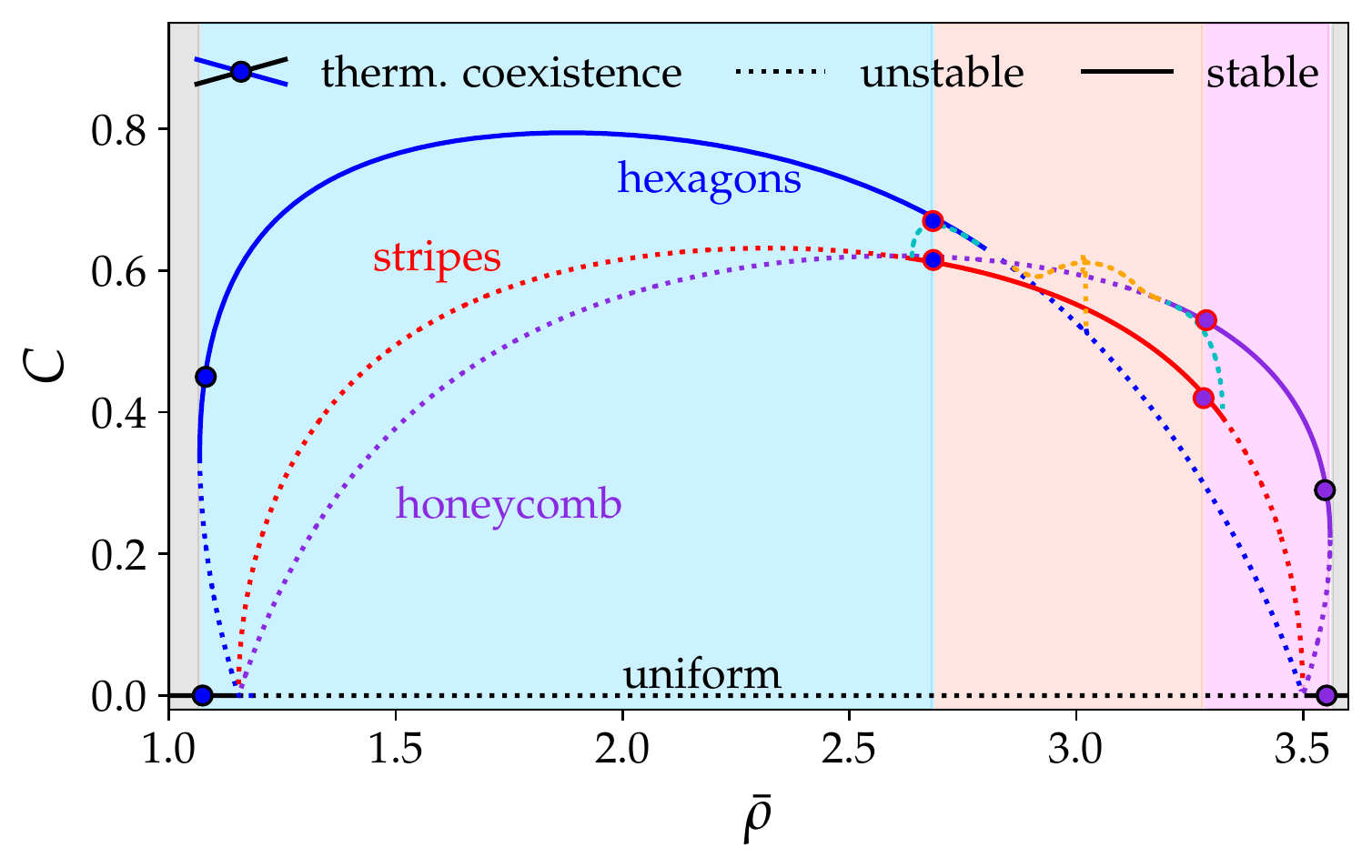} 
\caption{\label{fig:2D_bif_b_hon} Bifurcation diagram as in Fig.~\ref{fig:2D_bif_b} at smaller $b=0.1175$ (along the lower horizontal line in Fig.~\ref{fig:pt_change}). Domain size, line styles, solution measure and markers are as in Fig.~\ref{fig:2D_bif_b}. The additional dotted orange lines correspond to intermediate states between hexagons and honeycombs. The background shading locally matches the colors of the respective ground state. Thermodynamic coexistence of pairs of states is marked by filled circles: the two colors of filling and contour refer to the respective coexisting states.} 
\end{figure}

Figure~\ref{fig:2D_bif_b_hon} shows the bifurcation diagram of steady states at fixed $\alpha_u=-0.4$ and $b=0.1175$ (along the lower horizontal line in Fig.~\ref{fig:pt_change}). Then shadings highlight regions where the four different possible states are globally stable, i.e., correspond to the ground states.
There are four transitions between them, for each we indicate respective pairs of coexisting states that have identical chemical potential and grand potential density and here show nearly the same value of $\bar{\rho}$ with the difference being of order $10^{-2}$ or lower. Again, in consequence of the coexisting states we expect localized states to exist that could be calculated in larger domains. The coexistences indicate that all transitions are of first order. There are the transitions at low and high densities between the uniform state and hexagon and honeycomb patterns, respectively. Both are related to a combination of transcritical and saddle-node bifurcations. Furthermore, the $\bar\rho$-range of globally stable stripe patterns ends at respective transitions to hexagons and honeycombs. Starting at low $\bar\rho$, the uniform state is energetically favoured up to $\bar{\rho}=1.07$, followed by hexagons up to $\bar{\rho}=2.7$. Then, stripes have the lowest energy until $\bar{\rho}=3.28$ where honeycombs take over. Finally, at $\bar{\rho}=3.56$ the uniform state becomes again dominant.

\begin{figure}
\includegraphics[width=\linewidth]{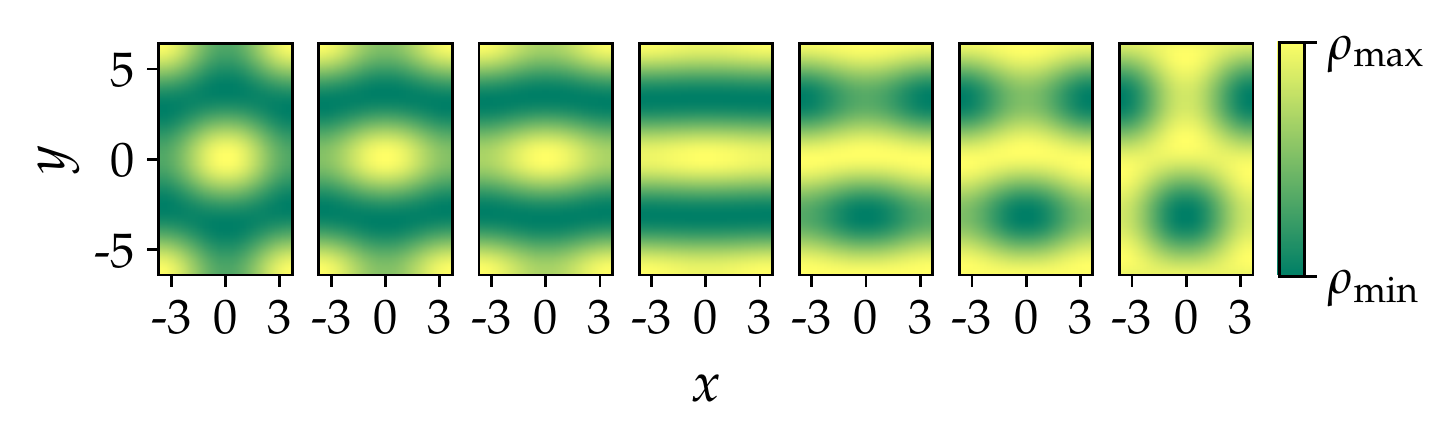} 
\caption{\label{fig:2D_transition} Examples of steady patterns illustrating the change from hexagons to stripes and further to honeycombs in Fig.~\ref{fig:2D_bif_b_hon} (for parameters see its caption). The sequence is as follows: From left to right there are a stable hexagon at $\bar{\rho}=2.75$, unstable intermediate states at $\bar{\rho}=2.7$ and $\bar{\rho}=2.69$, a stable stripe at $\bar{\rho}=2.67$, unstable intermediate states at $\bar{\rho}=3.27$ and $\bar{\rho}=3.25$, and a stable hexagon at $\bar{\rho}=3.19$.}

\end{figure}

The line connecting stable and unstable hexagons [honeycombs] is interrupted by a small gap, as numerically the corresponding pitchfork bifurcation becomes a perturbed pitchfork because the numerical grid does not fully reflect the symmetries of the involved patterns.
Typical density profiles from the branch of intermediate states between stripes and hexagons/honeycombs are given in Fig.~\ref{fig:2D_transition}. The sequence starts with a stable hexagon pattern (leftmost image), transforms into stripes (central image), before becoming a honeycomb (rightmost image). Close inspection  shows why it is necessary to use appropriate BC in the $y$-direction: while the change from hexagon to stripe works well with Neumann BC, the subsequent change from stripe to honeycomb requires periodic BC. Note that the reverse is true when starting from a periodic honeycomb pattern that is placed in such a way in the finite domain that Neumann BC apply. This issue may lead to bifurcations of plane-filling patterns being missed when considering finite domains. In consequence it may even result in stripe patterns that are shifted by half a period showing different stabilities. Therefore, bifurcation diagrams for finite domains have to be carefully interpreted.

There exist further intermediate states on connecting branches between hexagon and honeycomb branches (orange dotted lines in Fig.~\ref{fig:2D_bif_b_hon}). These unstable patterns include triangles and patchwork quilt patterns (as in Ref.~\onlinecite{Hoyl2007cup}, chapter~5.4). We do not exclude the possibility of further connecting branches between these states, but expect all of them to be unstable. Note, however, that in consequence of the various unstable states time evolutions may show intricate behaviour in the $\bar\rho-$range around $\bar\rho=3.0$ as unstable states may first attract and then repel the evolution, e.g., a time simulation started with perturbed triangle patterns, first approaches an unstable patchwork quilt pattern before converging to the globally stable stripe pattern (not shown).

\section{\label{sec:4}Summary \& Conclusions}

In the present work, we have investigated a PFC type model for BECs, employing arclength continuation techniques, stability analysis and one-mode approximations. The results have been presented in the form of bifurcation diagrams and phase diagrams. As a first step, we have approximated the nonlocal GPE in the vicinity of the phase transition by a PFC model. To do so we have proposed an alternative approach to Ref.~\onlinecite{HeBD2019pra} that aims at matching the dispersion relations of nonlocal and approximated local model at the onset of instability of the uniform state. The approach is general and does not depend on the specifics of the interatomic interactions (or nonlocal nonlinearity in case of optical systems), as long as the dispersion relation shows the expected maxon-roton form (monotonic small-scale instability).

We have then benchmarked the validity of the PFC model as an approximation for the behavior of the full nonlocal GPE (Eq.~\eqref{eq:gp}): The PFC model is able to correctly predict the onset of modulational stability and shows qualitative agreement with the GPE regarding the order of the phase transition. However,  there is a clear quantitative disagreement between the models when comparing finite-amplitude modulated steady states. This can be expected as we only expanded to fourth order.

Subsequently, we have presented bifurcation diagrams for spatially one- and two-dimensional systems, detailing possible steady states, coexistence regions and the existence of localized states. To the authors knowledge, localized states have not been discussed in depth in the context of BECs. Their analysis could turn out to be quite useful to experiments, as rich information may be extracted from individual images of localized states.

As expected, in the analyzed case no localized states have been encountered in a one-dimensional system with cubic nonlinearity.~\cite{Land1937zetf} However, we have found localized states in the two-dimensional system. Although they were found employing an approximative local model, we have conjectured that these states also persist in the full nonlocal model. This has further been substantiated by numerically showing their existence and stability for the full nonlocal equation.

Here, it has turned out to be numerically convenient to choose sufficiently low values of the interaction parameter $\alpha_u$, as this leads to a smaller spatial region of transition between two coexisting states. Nonetheless, we have to caution that the model is not suitable for excessively low values of $\alpha_u$, which result in unphysical negative densities. Note that we did not encounter slanted snaking of branches of localized states, which could have been expected in analogy to similar systems.~\cite{TARG2013pre} Most likely this is due to computational limitations that do not allow us to consider very large domains at interaction parameters sufficiently far from the critical point. 

In the final part we have considered how a phenomenological addition of a higher-order nonlinearity (akin to quantum fluctuations in dipolar BECs) changes the structure of the bifurcation diagram. We have found that such a nonlinearity stabilizes superfluid states with high densities and adds additional stable patterns. This is in close similarity to what has been recognized in dipolar BECs,~\cite{ZhMP2019prl,ZhPM2021pra,HSGB2021prr} thereby underpinning that, despite of our dramatically different model, the general behavior and topology persist.

In the present work we have been restricted to the study of spatially one- and two-dimensional systems. A remaining future challenge is to extend the results to three-dimensional systems. e.g., by adopting methods like the ones used in Ref.~\onlinecite{SAKR2016prl}.

In summary, we have presented an approximate mapping that permits one to employ standard bifurcation software and subsequently investigated bifurcation diagrams that include unstable states. In this way we have gained a deeper understanding of the specifics of the bifurcations present in the system. We believe that our results will remain qualitatively valid for other nonlocal GPE models and provide a reference case for their future bifurcation analysis. Note that the  presented methods can also be applied to nonlocal optical systems that display pattern formation.~\cite{SHAP2011prl,MPSK2016prl} Furthermore, the mapping could turn out to be useful for fast calculations of approximate excitation spectra as the matrix occurring in the Bogoliubov-de-Gennes equations that needs to be diagonalized is no longer full.~\cite{MMCP2013pra,GBHS2019nature}

\section*{Author Declarations}
The authors have no conflicts to disclose.

\begin{acknowledgments}
Discussions with Max Holl and Tobias Frohoff-H\"ulsmann are acknowledged. F.M. acknowledges stimulating discussions with Yongchang Zhang and Thomas Pohl. 
\end{acknowledgments}

\section*{Data Availability Statement}

The data that support the findings of this study are openly available on zenodo at http://doi.org/ 10.5281/zenodo.6594198, in Ref.~\onlinecite{SMGT2022data}.

\appendix 
\section{Localized State in the Nonlocal System}
\label{app:loc}
\begin{figure}
    \centering
    \includegraphics[width=0.9\textwidth]{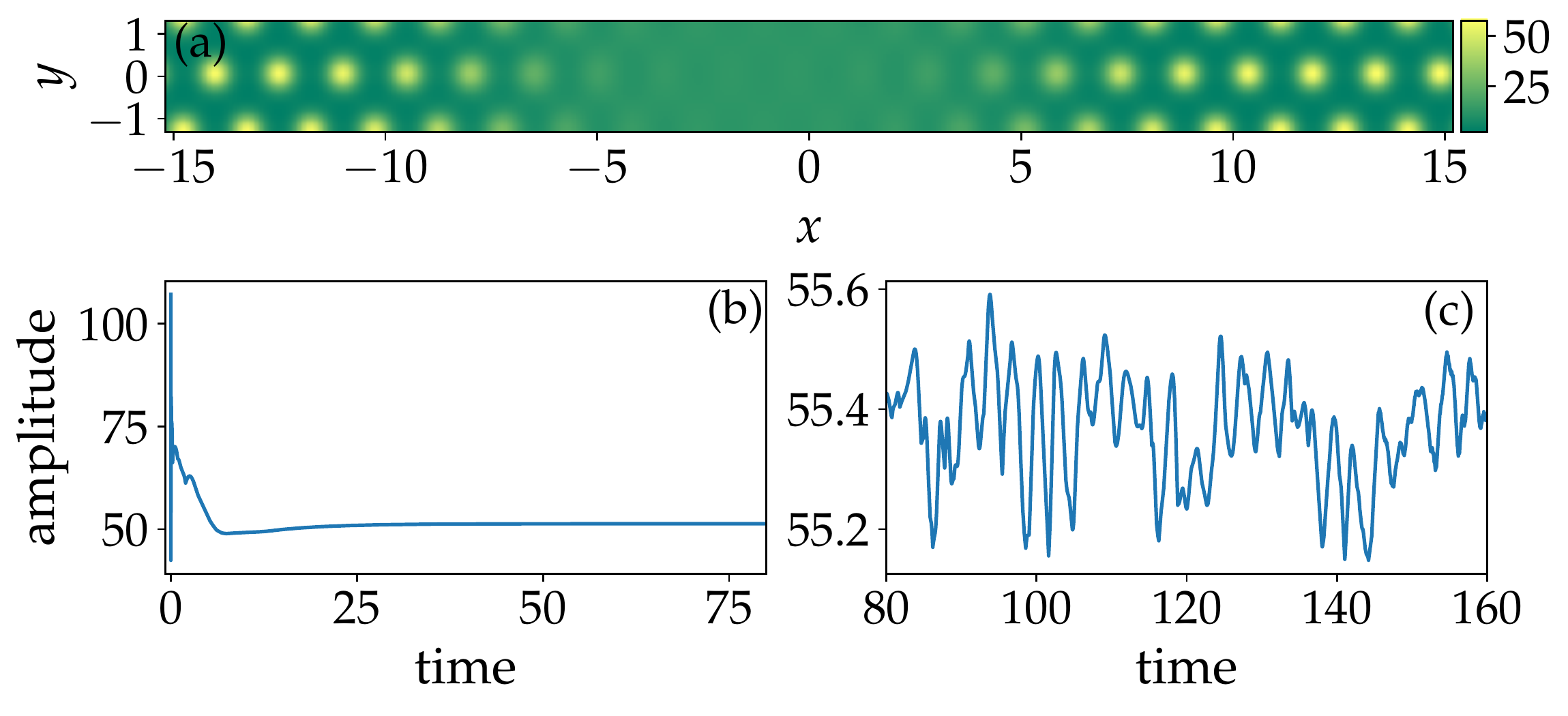}
    \caption{(a) Localized ground state for a  two-dimensional domain as obtained by complex time evolution of Eq.~\eqref{eq:gp}, i.e.\ the nonlocal GPE.
        (b) Relaxation of the amplitude during the complex time evolution. (c) Real-time evolution of the amplitude of previously relaxed state, indicating its robustness. The parameters are $a=0$ and $\bar{\rho}^\prime=12.625$ with a time step of $dt=10^{-3}$. The domain size is $L_x\times L_y$ with $L_x=48L_c^\prime/\sqrt{3}$ and $L_y=2L_c^\prime$ where $L_c^\prime=2\pi/kc^\prime$.
    }
    \label{fig:loc_state_time_simulation}
\end{figure}

To substantiate our hypothesis of the existence of localized states in the nonlocal system we perform a complex time evolution of Eq.~\eqref{eq:gp}, i.e.\ the nonlocal GPE, as described in Section~\ref{sec:2B_numerics}. In particular, we consider a two-dimensional domain for $a=0$ and $\bar{\rho}^\prime=12.625$ using a uniform state with randomly perturbed phase as initial condition.
This results in the localized state displayed in Fig.~\ref{fig:loc_state_time_simulation}~(a), which represents the ground state of the system. The relaxation of the amplitude during the complex time evolution is shown in Fig.~\ref{fig:loc_state_time_simulation}~(b). Finally, to probe the robustness of the result, we evolved this ground state using real-time evolution, i.e., employing a dissipative pseudo dynamics. The resulting small numerical fluctuations of the amplitude are illustrated in Fig.~\ref{fig:loc_state_time_simulation}~(c) clarifying that the pattern is preserved.
\section{Deriving the Approximation Parameters with Higher Nonlinearities\label{app:g2j+qmf}}

Starting from the dispersion relation in Eq.~\eqref{eq:disp_beyond}, we once again replace $\hat{R}(k)$ with the polynomial in Eq.~\eqref{eq:Rk} and demand that the requirements from Eqs.~\eqref{eq:req1}-\eqref{eq:req3} hold.

Given the critical values $k_c$ and $\bar{\rho}_c$ of $k$ and $\bar{\rho}$ we can now determine $g_0$, $g_2$ and $g_4$:
\begin{align}
g_0&=\frac{2\hat{R}(k_c)+4k_c\partial_{k}\hat{R}(k_c)+k_c^2\partial_{kk}\hat{R}(k_c)}{8}+\frac{3k_c^2}{8\bar{\rho}_c}-\frac{9b\sqrt{\bar{\rho_c}}}{8}\,,\\
g_2&=-\frac{1}{k_c^2}\left(2g_0+\frac{k_c^2}{4\bar{\rho}_c}+3b\sqrt{\bar{\rho_c}}\right)\,,\\
g_4&=\frac{1}{k_c^4}\left(g_0+\frac{3}{2}b\sqrt{\bar{\rho_c}}\right)\,.
\end{align}
$\partial_{k}\hat{R}$ and $\partial_{kk}\hat{R}$ are the first and second derivatives of $\hat{R}$ with respect to $k$.

Those $g_0$, $g_2$ and $g_4$ can now be used in the rescaling as before, with the addition of:
\begin{eqnarray}
 b \rightarrow b^\prime \frac{\alpha_u+1}{g_0}\sqrt{\frac{2}{-g_2}}\,.
\end{eqnarray}
\section{One-Mode Approximation\label{app:1mode}}
The one-mode approximation is used to get the minimum energy by inserting the ansatz
\begin{eqnarray}
   \rho=\bar{\rho}+\bar{\rho}A\sum_{i=1}^{3}\cos{(\bm{k}_i\bm{r})}
\end{eqnarray}
for hexagons [honeycombs] with $A>0$ [$A<0$] or
\begin{eqnarray}
 \rho=\bar{\rho}+\bar{\rho}A\cos{(\bm{k}_1\bm{r})}
\end{eqnarray}
for one dimensional solutions or stripes in two dimensions, into the energy $F$ given by Eq.~\eqref{equ:Ffunctional_b}. The lattice vectors $\bm{k}_i$ are  $\bm{k}_1 = (k,0)$, $\bm{k}_2 = (-k/2,k\sqrt{3}/2)$ and $\bm{k}_3 = (-k/2,-k\sqrt{3}/2)$. The amplitude of the pattern is $\bar{\rho}A$ and $A\ll1$.

We get results of the form
\begin{eqnarray}
 F-F_0= q_2A^2+ q_3A^3+ q_4A^4\,,\label{eq:1mode_app}
\end{eqnarray}
where $F_0$ is the uniform ground state energy at $A=0$. 

From there the energy is minimized first with respect to $k$, therefore $\partial_kF=0$ at $k=k_c$. Inserting $k_c$ into $F$ and neglecting all terms of order $O(A^5)$ and higher gives the results in Tab.~\ref{tab:one-mode_FA}.
\begin{table}[]
    \centering
    \begin{tabular}{c|c|c}
         & hexagons/honeycombs & 1D/stripes \\ \hline
         $q_2$ & $\frac{192\alpha_u\bar{\rho}^2 + 48\bar{\rho} + 288b\sqrt{\bar{\rho}}^5 - 3}{256}$ 
               & $\frac{ 64\alpha_u\bar{\rho}^2 + 16\bar{\rho} +  96b\sqrt{\bar{\rho}}^5 - 1}{256}$ \\ \hline
         $q_3$ & $\frac{-24\bar{\rho}+48b\sqrt{\bar{\rho}}^5+3}{256}$ 
               & $0$ \\ \hline
         $q_4$ & $\frac{240\bar{\rho} - 90b\sqrt{\bar{\rho}}^5 - 33}{1024}$ 
               & $\frac{  8\bar{\rho} -  3b\sqrt{\bar{\rho}}^5 -  1}{ 512}$ \\
    \end{tabular}
    \caption{Prefactors to the amplitudes in the energy equation (Eq.~\eqref{eq:1mode_app}) of the one mode approximation up to fourth order, depending only on the model parameters.}
    \label{tab:one-mode_FA}
\end{table}

In one dimension the sign of $q_4$ distinguishes between super- ($q_4>0$) and subcritical ($q_4<0$) pitchfork bifurcation. At $b=0$ we get $q_4>0$ for $\bar{\rho}>0.125$ meaning the bifurcation is supercritical as seen in the examples above. A subcritical pitchfork is not possible as $\bar{\rho}<0.125$ and therefore $\alpha_u<-1$ is not a valid parameter region in this model.

For $b\neq0$ we set $b$ so it matches the border of linear stability (see Eq.~\eqref{fig:lin_stab_b}) where the bifurcation emerges and get $ q_4 = \dfrac{64\alpha_u\bar{\rho}^2+272\bar{\rho}-33}{16384}$, meaning 
\begin{eqnarray}
 \frac{33}{64\bar{\rho}^2}-\frac{17}{4\bar{\rho}}\begin{cases}
    < \alpha_u \rightarrow \text{supercritical}\,,\\
    > \alpha_u \rightarrow \text{subcritical} \,.
\end{cases}
\end{eqnarray}
The former is always true for $0.125<\bar{\rho}<4.125$ as the model requires $\alpha_u>-1$. Beyond that subcritical bifurcations are possible.
\nocite{*}
\bibliography{bib}

\end{document}